\RequirePackage{fix-cm}
\documentclass[smallextended]{svjour3}       
\smartqed  
\usepackage{graphicx,color,soul}
\usepackage{lineno,bm}
\usepackage{algorithmicx}
\usepackage{algpseudocode}
\usepackage{algorithm}
\begin{document}

\title{The LBPM software package for simulating multiphase
flow on digital images of porous rocks 
}

\titlerunning{LBPM for multiphase flow}        

\author{James E. McClure$^1$         \and
        Zhe Li$^1$ \and 
        Mark Berrill$^2$ \and
        Thomas Ramstad$^3$ 
}


\institute{J. E. McClure, \email{mcclurej@vt.edu} \\
          1. Advanced Research Computing, Virginia Tech, Blacksburg, Virginia 24061, USA. \\
          2. Oak Ridge National Laboratory, Tennessee 37831, USA. \\
          3. Equinor ASA, Arkitekt Ebbells veg 10, Rotvoll, Trondheim, Norway. \\
}


\date{Received: date / Accepted: date}

\maketitle

\begin{abstract}
Direct pore scale simulations of two-fluid flow on digital rock images provide a promising tool to understand the role of surface wetting phenomena on flow and transport in geologic reservoirs. We present computational protocols that mimic conventional special core analysis laboratory (SCAL) experiments, which are implemented within the open source LBPM software package. Protocols are described to simulate unsteady displacement, steady-state flow at fixed saturation, and to mimic centrifuge experiments. These methods can be used to infer relative permeability and capillary curves, and otherwise understand two-fluid flow behavior based on first principles. Morphological tools are applied to assess image resolution,
establish initial conditions, and instantiate surface wetting maps based on the distribution of fluids.
Internal analysis tools are described that measure essential aspects of two-fluid flow, 
including fluid connectivity and surface measures, which are used to track transient aspects of the
flow behavior as they occur during simulation. Computationally efficient workflows are developed by
combining these components with a two-fluid lattice Boltzmann model to define hybrid methods that
can accelerate computations by using morphological tools to incrementally evolve the pore-scale fluid distribution. We show that the described methods can be applied to recover expected trends due to the surface wetting properties based on flow simulation in Benntheimer sandstone. 

\end{abstract}

\section{Introduction}
\label{intro}

 LBPM (Lattice Boltzmann Methods for Porous Media) is an open source software framework designed to model flow processes based on digital rock physics, and is freely available through the {\it Open Porous Media} project \cite{OPM_2011}. Digital rock physics refers to a growing class of methods that leverage microscopic data sources to obtain insight into the physical behavior of fluids in rock and other porous materials \cite{ISI:000314105900011}. Rock micro-structure is highly complex, as are the interactions between {\em in-situ} fluid-fluid and fluid-rock interfaces. 
As data from X-ray micro-computed tomography ($\mu$CT) and other image acquisition technologies become more widely available, approaches to directly tie these data sources to reservoir-scale flow and transport processes become more generally useful. Direct simulation of multiphase flow at the pore scale is gaining ground as a viable alternative to expensive experimental approaches. Key technological advances have been responsible for the rise of these techniques as a viable means of advancing knowledge for these systems:
(1) the capability to experimentally image two-fluid flow processes down to micron resolution; and
(2) advances in computing that allow for direct simulation of two-fluid flow in complex geometries. Pore-scale information not only provides a way to study mechanisms for multiphase flow, 
but also to study larger scale phenomena based on direct upscaling. Key parameters for reservoir 
simulation are the relative permeability and capillary pressure \cite{bear1988dynamics,doi:10.1029/WR012i003p00513}. 
For practical reasons subsurface transport phenomena must be described based on averaged flow properties on large length scales, which depend on the distribution and dynamics of fluids within the pore space. 

Surface wetting phenomena are known to influence reservoir-scale flow and transport. Within the pore structure, fluids will tend to assume an arrangement that minimizes the potential energy. Capillary forces usually dominate at the pore-scale, and surface energies represent a significant contribution to the overall potential energy of the system.  Due to the influence of surface forces on the pore-scale distribution of fluids, the relative permeability and capillary pressure relations are sensitive to the
effects of wetting. Various special core analysis laboratory (SCAL) 
workflows have been developed to characterize the associated behaviors.
Direct simulation offers new opportunities to advance understanding for these
complex systems by providing a mechanism to directly control surface wetting properties and assess their effects. Lattice Boltzmann methods can be constructed to operate based on minimal assumptions, and are well-suited toward modeling fluids within complex materials without relying on geometric simplifications. Multiphase simulations directly resolve the evolution of the pore-scale fluid distribution, pressure and velocity 
fields, and can be used to assess pore-scale flow mechanisms as well as 
evaluate larger-scale averaged behavior. It is straightforward to manipulate 
surface wetting properties within a simulation, meaning that many scenarios can
be considered, including cases that would be intractable in a traditional experimental setting. 

We present simulation protocols based on two-fluid lattice Boltzmann methods, focusing in particular on wetting phenomena. Efficient and fast set-ups are developed as  computational analogs for existing SCAL workflows. Strategies to assess reservoir-scale sensitivities based on local physical properties are developed. The associated direct simulations can be used to reduce the uncertainty associated with lab data and improve general understanding on wetting phenomena in
complex systems. Essential effects due to the capillary number, viscosity ratio and flow history can be controlled explicitly and incorporated into simulation workflows.
In the following sections, we present basic background information on multiphase flow 
and describe methods that have been implemented to model complex wetting phenomena
within the context of digital rocks. We then describe simulation protocols that are implemented within LBPM, to facilitate efficient simulation studies \cite{opm_lbpm}. Results are presented for two-fluid flow within a well-resolved 
$\mu$CT image of a Bentheimer sandstone \cite{Dalton_2019}, which is a standard rock type for experimental benchmark studies. Both unsteady displacement and steady-state flow simulations are conducted, and results are demonstrated to recover expected behaviors.

\section{Background}

\begin{figure}
\includegraphics[width=0.5\textwidth]{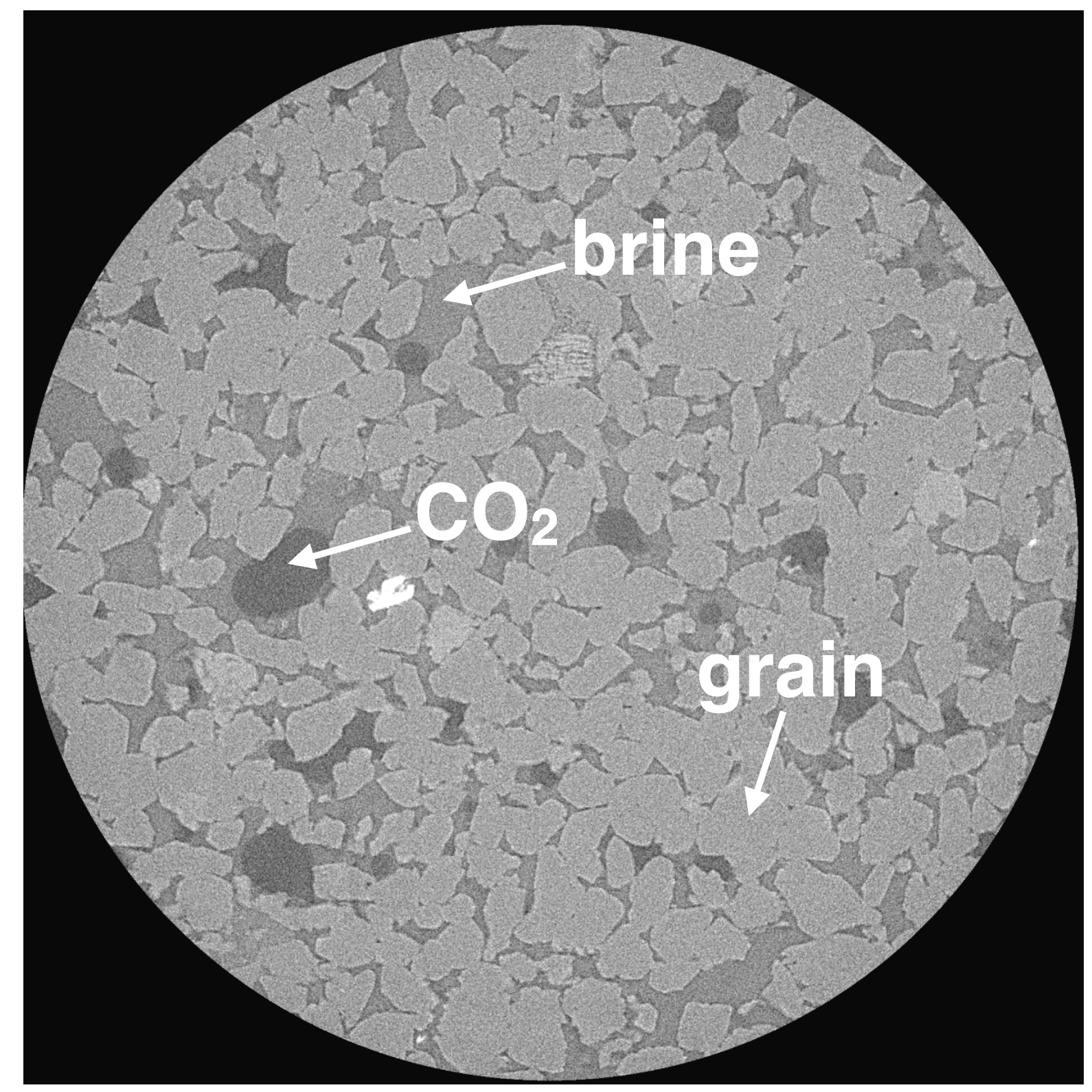}
\caption{ Slice of experimental $\mu$CT image showing
{\em in situ} distribution of brine and $\mbox{CO}_2$ within
 Bentheimer sandstone \cite{Dalton_2019}.
}  
\label{fig:uCT}
\end{figure}

Direct pore-scale simulation of fluid flow is possible based on the availability of $\mu$CT
data that resolves the actual rock micro-structure. Current capabilities allow for 3D imaging
that resolves both rock structure and {\em in situ} fluid distributions, as shown in 
Fig. \ref{fig:uCT}. Volumetric data sets that are reconstructed from experimental observations must be post-processed to remove noise and other imaging artifacts so that the spatial distribution of materials within the sample can be reliably identified. The associated segmentation workflows have been treated in detail by other authors (e.g. \cite{Wildenschild_Sheppard_13}). Simulation results depend on sufficiently well-resolved images and high-quality segmentation. While LBPM does include some data pre-processing tools for experimental data, the two-phase simulator relies on segmented data for input. It is possible to ingest information about both fluid and solid materials based on experimental images. 

Flow through porous media is described based on a hierarchy of approximations that are rooted in the work of Darcy. Darcy's experiments related the pressure drop across a single fluid system to the associated flow rate \cite{darcy1856fontaines}. When multiple fluids are present, each fluid occupies a portion of the porespace, competing with other fluids for the dominant flow pathways. Multiphase extensions of Darcy's law are inspired by a conceptual model in which each fluid is arranged to form a connected pathway through the material, which supports flow. The extended version of 
Darcy's law is stated as follows:
\begin{equation}
    \mathbf{u_i} =  \frac{k_i}{\mu_i} \big(\rho \mathbf{g} - \nabla p_i \big), \;
    \label{eq:kr-sw-original}
\end{equation}
where $\mathbf{u_i}$ is the flow velocity for fluid $i\in\{a,b\}$ and $\mu_i$ is the associated dynamic viscosity. The driving forces are the external body force $\mathbf{g}$ and the pressure gradient $\nabla p_i$. The effective permeability $k_i$ accounts for the influence of the fluid structure on the rate of energy dissipation for the flow process. Since the effect of the material is accounted for based on the absolute permeability $K$, relative permeability is often defined based on the ratio:
\begin{equation}
k^r_{i} = \frac{k_i}{K}\;.
\end{equation}
The effective permeability depends on the particular arrangement of each fluid
within the material, which changes based on the fluid volume fraction and flow history \cite{bear1988dynamics}.
Traditionally the relative permeability is taken to depend on the fluid saturation $S_w$,
which accounts for first-order effects due to the configuration of fluids. It 
is well-known that the dependence $k^r_i(S_w)$ is non-functional for typical materials \cite{doi:10.1029/WR012i003p00513}.
Surface forces acting at the pore-scale can also strongly influence the fluid configuration,
which includes capillary forces and effects due to the wetting properties of solid materials \cite{Jadhunandan_Morrow_1995}. A key opportunity for pore-scale simulation studies is to resolve these dependencies so that the consequences for larger scale transport can be understood \cite{Ramstad_Idowu_12}.
Along with the relative permeability, a constitutive relationship for the
capillary pressure is included in reservoir models. From the microscopic perspective,
capillary pressure $p_c$ arises due to the fluid-fluid meniscus curvature and interfacial tension
\cite{Adamson_Gast_97}. The capillary force causes a difference between the fluid pressures, 
$p_c = p_a-p_b$. Since the fluid pressures are needed based on Eq. \ref{eq:kr-sw-original},
the capillary pressure is needed to construct a closed reservoir-scale model. Like the relative permeability, $p_c$
is traditionally assumed to depend on $S_w$, neglecting transient effects and the role of fluid
history in determining the fluid configuration. Uncertainties associated with this approximation
are usually presumed to be less significant than other model uncertainties, such as length-scale
heterogeneity within the reservoir. 

The traditional multiphase extension of Darcy's law is a source of many challenges and contradictions \cite{doi:10.1029/98RG00878}. While Eq. \ref{eq:kr-sw-original} states a linear relationship between the potential gradient and flow rate, the effective permeability is linked with a variety of non-linear behaviors. Most typically these behaviors are characterized
in terms of the predominant non-dimensional quantities. First is the capillary number ($\mbox{Ca}$), 
which accounts for the balance of capillary and viscous forces in the system:
\begin{equation}
    \mbox{Ca} = \frac{\mu_i |\mathbf{u}_i|}{\sigma_{ab}}\;,
    \label{eq:Ca}
\end{equation}
where $\mu_i$ is the dynamic viscosity for fluid $i$, $\mathbf{u}_i$ is the flow velocity, and 
$\sigma_{ab}$ is the fluid-fluid interfacial tension. The viscosity ratio ($\mbox{M}$) can also
play a significant role, defined as:
\begin{equation}
    \mbox{M} = \frac{\mu_a}{\mu_b} \;.
    \label{eq:M}
\end{equation}
Relative permeability depends non-linearly on the viscosity ratio and capillary number, particularly for  $\mbox{Ca} \ge 10^{-5}$ \cite{Lenormand1989}. The relative permeability is known to be process-dependent, and different curves are obtained along drainage and imbibition due to the fact that the structural properties of the fluids depend on the system history. Structural effects are centrally important for multiphase flows, and have a deterministic impact on macroscopic flow behavior such as 
capillary pressure and relative permeability \cite{Erpelding_Sinha_etal_13}. Indeed, the role of the
essential non-dimensional quantities is in many cases to alter the configuration of fluids.
When the difference between the fluid densities becomes large, the Bond number ($\mbox{Bo}$) can also become important:
\begin{equation}
    \mbox{Bo} = \frac{(\rho_a -\rho_b) |\mathbf{g}|D^2}{\sigma_{ab}} \;,
    \label{eq:Bond}
\end{equation}
where $D$ is a characteristic length scale for the system (e.g. Sauter mean diamter)
and $\mathbf{g}$ is the body force acting on the system. The role of intertial 
effects can be included based on the Ohnesorge number:
\begin{equation}
    \mbox{Oh} = \frac{\mu_i}{\sqrt{\rho_i \sigma_{ab} D} }\;.
    \label{eq:Ohnesorge}
\end{equation}
The Ohnesorge number accounts for the effect of the Reynolds number in a system
where capillary forces are also present. For two-fluid flows Reynolds number effects 
are generally less important than capillary number effects due to the dominance
of capillary forces at small length scales, which inhibits the formation of eddies
that can play a strong role in single fluid flows \cite{PhysRevE.87.033012,McClure_2010}. 
The Ohnesorge number can nevertheless be important for certain applications in multiphase flow, 
such as droplet formation, which occur on a fast timescale.

\begin{figure}
\centering
\includegraphics[width=0.7\textwidth]{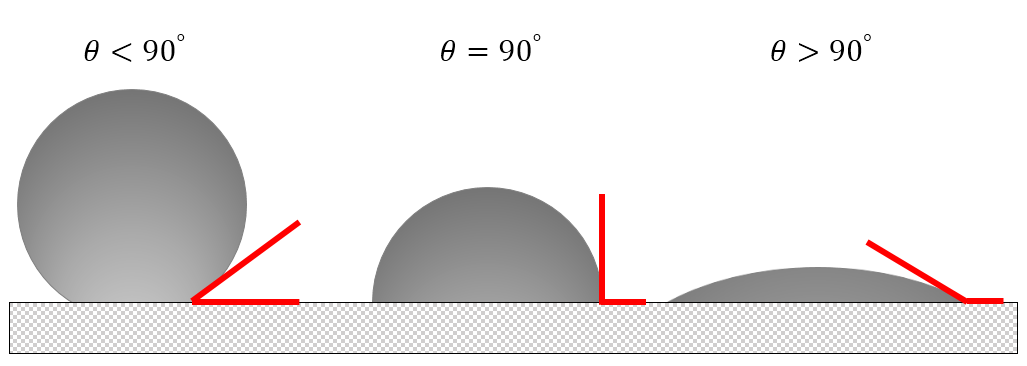}
\caption{ Example of contact angle where a liquid droplet is situated on a solid surface.
}  
\label{fig:sketch_contact_angle}
\end{figure}

In real systems surface wetting properties are often heterogeneous, and the associated
fluid-solid interactions influence the pore-scale distribution of fluids as well as the 
flow behavior. A straightforward non-dimensional representation of surface wetting can be constructed
from the three interfacial energies:
\begin{equation}
    \mbox{W} = \frac{\sigma_{as}- \sigma_{bs}}{\sigma_{ab}} \;,
    \label{eq:W}
\end{equation}
which relates to the equilibrium contact angle based on 
Young's equation for the contact line: \cite{Adamson_Gast_97}
\begin{equation}
\cos \theta_{eq} = \frac{\sigma_{as}-\sigma_{bs}}{\sigma_{ab}}\;.
\end{equation}
An illustrative example of static contact angle is shown in Fig.\ref{fig:sketch_contact_angle}.
Since the fluid-solid interfacial energies are a local property along the
rock surface, a wide range of complex behaviors are possible based on surface heterogeneity. 
Surface properties impact the pore-scale distribution of fluids, and consequently
can have a significant impact on the resulting flow behavior. Sensitivities with respect to wetting behavior are of particular interest for reservoir modeling, since these effects will influence the relative permeability even at the low capillary numbers associated with reservoir-scale flow processes. 

Taking note of the important dimensionless quantities, Eq. \ref{eq:kr-sw-original} can be put into a non-dimensional form by dividing by $\sigma_{ab}$ and re-arranged as:
\begin{equation}
    \frac{\mu_i \mathbf{u_i}}{\sigma_{ab}} =  \frac{k_i}{D^2}
    \frac{\big(\rho \mathbf{g} - \nabla p_i \big)D^2}{\sigma_{ab}}, \;
    \label{eq:kr-sw}
\end{equation}
where three non-dimensional terms can be easily identified, with the non-dimensional form for the effective permeability being $k_i / D^2$; the left-hand side
is directly linked to the capillary number; the driving forces are 
expressed in a way that is very similar to the Bond number, but accounting 
for forces acting on the system in a more general way. The non-dimensional
form is particularly useful to interpret simulation results, since effects
that are solely due to the choice of units are removed. 

Pore-scale simulation provides a powerful framework to understand and characterize the
uncertainties related to the relative permeability because these uncertainties originate from
pore-scale. The distribution of fluids within rock and their associated flow behaviors can 
be determined from direct simulation based on knowledge of the rock micro-structure. Experimental
techniques such as $\mu$CT can provide the requisite data for many geologic materials of interest. Digital rock workflows are constructed as a way to add value to experimental workflows
by generating additional information about the physical behaviors of the observed system.
Computational experiments can be performed to explore a wide-range of scenarios, considering 
situations that would often be impossible or impractical in an experimental setting.
For multiphase flows, much of the advantage comes from being able to simulate flow processes
by matching the various dimensionless numbers that control the physical behavior. Immiscible fluid flow in the subsurface is often dominated by capillary forces at small length scales, and viscous forces at large length scales. The capillary number, viscosity ratio and wettability account for the first order effects of these forces on the relative permeability, and are important to capture the flow regime. 

LBMs are attractive for digital rock physics studies because they can
model a wide range of physics, and can be applied to match critical dimensionless numbers
and account for their influence in natural and engineered systems.
In this work we rely on a color LBM, which is widely used to model immiscible fluid flow
through complex geometries. In typical digital rock images, the color LBM tends to be stable for a wide range capillary numbers, $10^{-6} < \mbox{Ca} < 1$, and is capable of modeling viscosity ratio $0.01<\mbox{M}<100$, with a similar range of stability for the density ratio. Since the LBM is a mesoscopic method, local surface wetting properties can be set based on simple rules that are able to account for both equilibrium and non-equilibrium effects \cite{Latva_Kokko_Rothman_05b,Latva_Kokko_Rothman_07}. Additional details on the method and its implementation within LBPM are provided in the sections that follow.

\section{Methods}
\label{sec:methods}

This work describes computational protocols to estimate relative permeability and capillary pressure based on digital rock images, with a particular focus on the role of wetting. 
Digital rock approaches provide a particularly powerful tool for understanding the role played by material structure in transport\cite{ISI:A1988N050200002,ISI:A1990DD11300001,ISI:000180779800003}.
The increased availability of 3D data from sources
such as X-ray  micro-computed tomography ($\mu$CT) is able to support a wide range of computational studies that rely on realistic rock micro-structure and direct physical simulation that rely on first-principles modeling \cite{Schlueter_Sheppard_etal_14,ISI:000082526700007,ISI:000089336000016}.
Digital imaging technologies are able to generate larger data volumes each year. 
However, our ability to extract useful information from these data depends to 
a large extent on whether or not $\mu$CT technologies are applied in a way
that allows us to measure important quantities reliably and accurately. 
Accurate and reliable measurements require that the pore-space is sufficiently 
well-resolved to measure quantities of interest. The generation of digital rock
images itself relies on non-trivial computational pipelines for image reconstruction and segmentation \cite{Wildenschild_Sheppard_13}. It is understood that
high-quality experimental data is an essential starting point for simulation workflows.  

LBMs are a computationally-efficient approach to model fluid flows in complex geometries and can be constructed to model a wide range of physics \cite{ISI:000382680900002,ISI:000341218200003,Joekar-Niasar_vanDijke_etal_12,Geller_Krafczyk_etal_06}. LBMs have been widely applied to study two-fluid flow in porous media, with efforts focusing on modeling fluid displacement \cite{ISI:000344068000012,ISI:000285177400003}, viscous coupling \cite{ISI:000246548700006,ISI:000269685800006}, and the development of workflows to recover relative permeability and capillary pressure \cite{ISI:000450094200004,ISI:000452345000009,Ramstad_Idowu_12,Thomas_SCA_2019,ISI:000285177400003}.
The use of lattice Boltzmann methods to simulate flow and transport in $\mu$CT images and estimate relative permeability measurement is well-known \cite{ISI:000449669500002,ISI:000432555400012,ISI:000405674300013,ISI:000404309200006,ISI:000394568900021,ISI:000388856300003,ISI:000383299300006,ISI:000383299300013,ISI:000321725100003,ISI:000307391900003,ISI:000291065000013,ISI:000275765200034,ISI:000270378600021,Hussain_Pinczewski_14,Landry_Karpyn_etal_14}. The simulations required to measure relative permeability are considerably more expensive than those required to measure the permeability. There are several reasons why this is true. First, multiple simulations are
required to measure the relative permeability over a range of fluid saturations. Second, multiphase simulations generally reach steady state
less quickly than single phase simulations. In a multiphase simulation, the approach to steady state is limited in part by the interfacial dynamics. 
Since interfacial dynamics are very slow compared to the pressure dynamics, a large number of timesteps are required to simulate multiphase flow processes.
For this type of problem, LBM is often the method of 
choice due to the complex structure of rocks and the computational costs associated with simulation in large digital images \cite{Navaraez_Zauner_etal_10,Namgyun_10,Maeir_Bernard_10,Toelke_Freudiger_etal_06,Ahrenholz_Toelke_etal_08}. The study of time-dependent problems is particularly important, since $\mathcal{O(}N)$ methods (e.g. multi-grid) are disadvantaged for unsteady flows \cite{Tolke_Krafczyk_etal_02,Geller_Krafczyk_etal_06}. Multiphase flows with moving interfaces fall into this category. The number of timesteps required to simulate displacement processes is dominated by the speed that the interfaces move through the system; creeping flows at low capillary numbers are associated with particularly large computational costs. Hybrid schemes are attractive because they provide a way to move the interfaces more rapidly so that results can be obtained at lower computational cost.
We describe the methodology used by LBPM to model both steady and unsteady fluid flows
that is applicable to the study of general wetting behaviors. We demonstrate how LBPM
computational protocols can be applied to directly assess capillary pressure and relative permeability based on 3D $\mu$CT images. 

\subsection{Color Lattice-Boltzmann Model}
\label{sec:LBM}

The color lattice-Boltzmann model is well-established for modeling two-fluid flow in porous media \cite{Ahrenholz_Toelke_etal_08,Latva_Kokko_Rothman_05b}. The scheme is defined by a set of three lattice Boltzmann equations (LBEs). The LBEs are defined based on a quadrature scheme to discretize the velocity space in the continuum Boltzmann equation. For the popular D3Q19 velocity set used in this work, the discrete velocities are given by 
\begin{equation}
 \xi_q = \left
  \{ \begin{array}{ll}
    \{ 0,0,0\}^T & \mbox{for $q=0$} \\
    \{ \pm 1,0,0\}^T,  & \mbox{for $q=1,2$} \\
    \{ 0,\pm 1,0\}^T,   &\mbox{for $q=3,4$} \\
    \{ 0,0,\pm 1\}^T  & \mbox{for $q=5,6$} \\
    \{ \pm 1,\pm 1,0\}^T, & \mbox{for $q=7,8,9,10$} \\
    \{ \pm 1,0,\pm 1\}^T, & \mbox{for $q=11,12,13,14$} \\
    \{ 0,\pm 1,\pm 1\}^T   & \mbox{for $q=15,16,17,18$}\;. \\
\end{array} \right.
\label{eq:d3q19}
\end{equation}
The first two LBEs model the mass transport behavior for each of two components in the system based on the number density for each component, $N_a$ and $N_b$ \cite{McClure_Prins_etal_14}.
The D3Q7 velocity set is sufficient to recover mass transport, which corresponds to 
$q=0,1,\ldots,6$ in Eq. \ref{eq:d3q19}. Two sets of distributions evolve based on the
following rules
\begin{eqnarray}
A_q(\bm{x} + \bm{\xi}_q \delta t, t+\delta t) &=& w_q N_a \Big[1 + \frac{\bm{u} \cdot \bm{\xi}_q}{c_s^2} 
+ \beta  \frac{N_b}{N_a+N_b} \bm{n} \cdot \bm{\xi}_q\Big] \;
\label{eq:LBE-mass-a}
\\
B_q(\bm{x} + \bm{\xi}_q \delta t, t+\delta t) &=& 
w_q N_b \Big[1 + \frac{\bm{u} \cdot \bm{\xi}_q}{c_s^2}
- \beta  \frac{N_a}{N_a+N_b} \bm{n} \cdot \bm{\xi}_q\Big]\;, 
\label{eq:LBE-mass-b}
\end{eqnarray}
where for the D3Q7 lattice, the speed of sound is $c_s=\sqrt{2}/3$, the weights are $w_0=1/3$ and $w_{1,\mbox{\ldots},6} = 1/9$, and $\beta$ controls the thickness of the phase interface region. The mass transport 
equations are constructed to minimize the diffusion of mass against the direction of the color gradient, which is given by:
\begin{equation}
\textbf{C}=\nabla \phi\;,
\end{equation}
where the phase indicator field $\phi$ is defined based on the number density for the two fluids:
\begin{equation}
\phi = \frac{N_a-N_b}{N_a+N_b}\;.
\end{equation}
The unit normal vector of the color gradient $\bm{n} = (n_x,n_y,n_z)$ is 
calculated as:
\begin{equation}
\bm{n} = \frac{\textbf{C}}{|\textbf{C}|}\;.
\end{equation}
The mass transport LBEs 
are coupled to a momentum transport LBE based on the flow velocity $\bm{u}$. The
D3Q19 quadrature rule is used for the momentum transport LBE, which relies on a multi-relaxation time (MRT) collision model,
\begin{equation}
f_q(\bm{x}_i + \bm{\xi}_q \delta t,t + \delta t) - f_q(\bm{x}_i,t) = \sum^{Q-1}_{k=0} M^{-1}_{qk} S_{kk} (m_k^{eq}-m_k) + t_q \bm{\xi}_q \cdot \frac{\bm{F}}{c_s^2} \;,
\end{equation}
where for the D3Q19 lattice, the speed of sound is $c_s=1/\sqrt{3}$, and the weights are $t_0=1/3$, $t_{1,\mbox{\ldots},6} = 1/18$, and $t_{7,\mbox{\ldots},18} = 1/36$. The coefficient matrix $M_{qk}$ and it's inverse $M^{-1}_{qk}$ are determined based on the work of d'Humi\'{e}res et al., and the diagonal relaxation matrix $S_{kk}$ is determined based on the kinematic viscosity \cite{dHumieres_Ginzburg_etal_2002,dHumieres_Ginzburg_2003}. A constant external body force $\bm{F}$ is imposed to drive the flow \cite{Lallemand_Luo_00}. The moments $m_k$,  for $k=0,1,\ldots,18$, are orthogonal linear combinations of the distributions $f_q$, for $q=0,1,\ldots,18$,
\begin{equation}
m_k = \sum_{q=0}^{18} M_{qk} f_q,
\end{equation}
For the color model, the non-zero equilibrium moments that incorporate the fluid-fluid interfacial tension are:
\begin{eqnarray}
m_1^{eq} &=& (j_x^2+j_y^2+j_z^2) - \alpha |\textbf{C}|, \\
m_9^{eq} &=& (2j_x^2-j_y^2-j_z^2)+ \alpha \frac{|\textbf{C}|}{2}(2n_x^2-n_y^2-n_z^2), \\
m_{11}^{eq} &=& (j_y^2-j_z^2) + \alpha \frac{|\textbf{C}|}{2}(n_y^2-n_z^2), \\
m_{13}^{eq} &=& j_x j_y + \alpha \frac{|\textbf{C}|}{2} n_x n_y\;, \\
m_{14}^{eq} &=& j_y j_z + \alpha \frac{|\textbf{C}|}{2} n_y n_z\;, \\
m_{15}^{eq} &=& j_x j_z + \alpha \frac{|\textbf{C}|}{2} n_x n_z\;, 
\end{eqnarray}
where the interfacial tension between fluids is $\sigma_{ab} = 6\alpha$.

\subsection{Modeling Wetting with Lattice Boltzmann models}
\label{sec:wetting}

An important application area for digital rock physics is to understand the 
role of wetting in two-fluid flows. The wetting condition
influences reservoir-scale transport behavior based on the sensitivity of the relative
permeability and capillary pressure \cite{Graue_Ferno_12}. It is also evident that the local
wetting state is subject to significant spatial variability in typical geological reservoirs
\cite{Saraji_Goual_etal_13,Matthew_Bijelic_etal_14}. Local wetting properties along the grain surface can vary due to the mineral type, fluid history, surface roughness and other factors \cite{Rucker_2020}. 
In designing simulation tools it is therefore important to allow for flexible assignment of
wetting conditions so that a wide range of scenarios can be simulated and studied. The implementation
within LBPM allows users to specify the wetting map in the form of a 3D digital rock image so that
generic spatial wetting distributions can be assigned. The wetting map assigns
a label to each solid voxel, and the associated wetting affinity is specified for each label.  The labels can thereby be assigned externally to LBPM so that customized wetting maps can be generated by the user.

Approaches to model basic wetting behaviors in porous media have been previously advanced using lattice Boltzmann methods \cite{ISI:000428474500035,ISI:000337672900004,Thomas_SCA_2019}. 
A number of schemes have been developed to model the effects of wetting
\cite{Huang_Thorne_etal_07,Wiklund_Lindstrom_etal_11,Schmieschek_Harting_11,Wolf_dosSantos_etal_09,Lee_Lin_08,Benzi_Biferale_etal_06a,Benzi_Biferale_etal_06c}.
From the strictly physical standpoint, wetting and spreading involve a wide range of complex behaviors, many of which are not well-understood from the continuum perspective
\cite{Pomeau_02,Dhori_Slattery_97,Shikhmurzaev_97,Brochardwyart_DeGennes_92,Seppecher_96}.
Mesoscopic methods (such as LBM) are very attractive within this context, since quasi-molecular 
rules can be assigned without relying on continuum-scale closure relationships. This provides
an effective way to generate strategies to model general behavior at the contact line 
\cite{Pooley_Kusumaatmaja_etal_08,Kawasaki_Onishi_etal_08,Latva_Kokko_Rothman_07}. It is important
to note the many challenges associated with the construction of such models: films,
moving contact lines, entrainment, contact line pinning, surface roughness and resolution pose thorny issues for numerical models \cite{Zhao13799}. While LBMs do provide
some basis to study these effects from first principles, our focus here is to consider the
cumulative impact of wetting on reservoir-scale constitutive models such as relative
permeability and capillary pressure. The predominant flow mechanisms are due
to the arrangement of fluids within complex solid mateirals. Within this context, LBMs 
provide excellent capabilities and associated sensitivities for wetting behavior.

\begin{figure}
\includegraphics[width=1.0\textwidth]{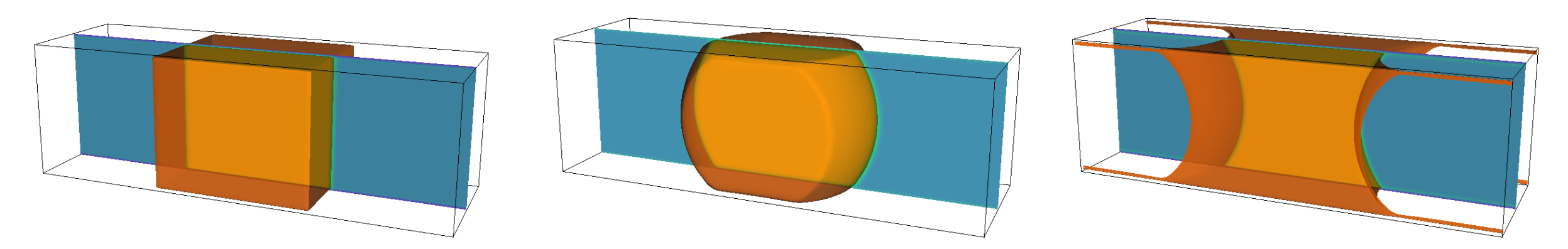}
\caption{Illustration of an oil bubble trapped in a square tube with solid surface being neutral-wet (left), water-wet (middle), and oil-water (right) condition. The phase indicator field at the central slice is also presented. 
}  
\label{fig:bubble-in-tube-demo}
\end{figure}

In the color LBM, the wetting condition can be defined from a scalar affinity value. It has been demonstrated that the affinity value is equivalent to a pseudo-phase field, $\phi_s$, which ranges from $-1.0$ to 1.0. This simple method provides a natural mechanism to assign the surface energy within pore-scale models, recovering expected contact line behavior both at equilibrium and for moving contact lines \cite{Latva_Kokko_Rothman_05a}. For example, Fig. \ref{fig:bubble-in-tube-demo} illustrates that a LB simulated oil bubble in water is trapped in a square tube at different wetting conditions. On a plain surface where there is a well-defined contact line among the solid, the equilibrium wetting condition is characterized by the equilibrium contact angle $\theta_{eq}$, which is linked to $\phi_s$ by:
\begin{equation}
    \cos \theta_{eq} = \phi_s.
    \label{eq:phi_s}
\end{equation}
We follow the commonly used terminology to refer a local wettability of $\phi_s(\bm{x})= 1.0$ to strongly water-wet, and $\phi_s(\bm{x})=-1.0$ to strongly oil-wet.

\begin{figure}
  \includegraphics[width=0.8\textwidth]{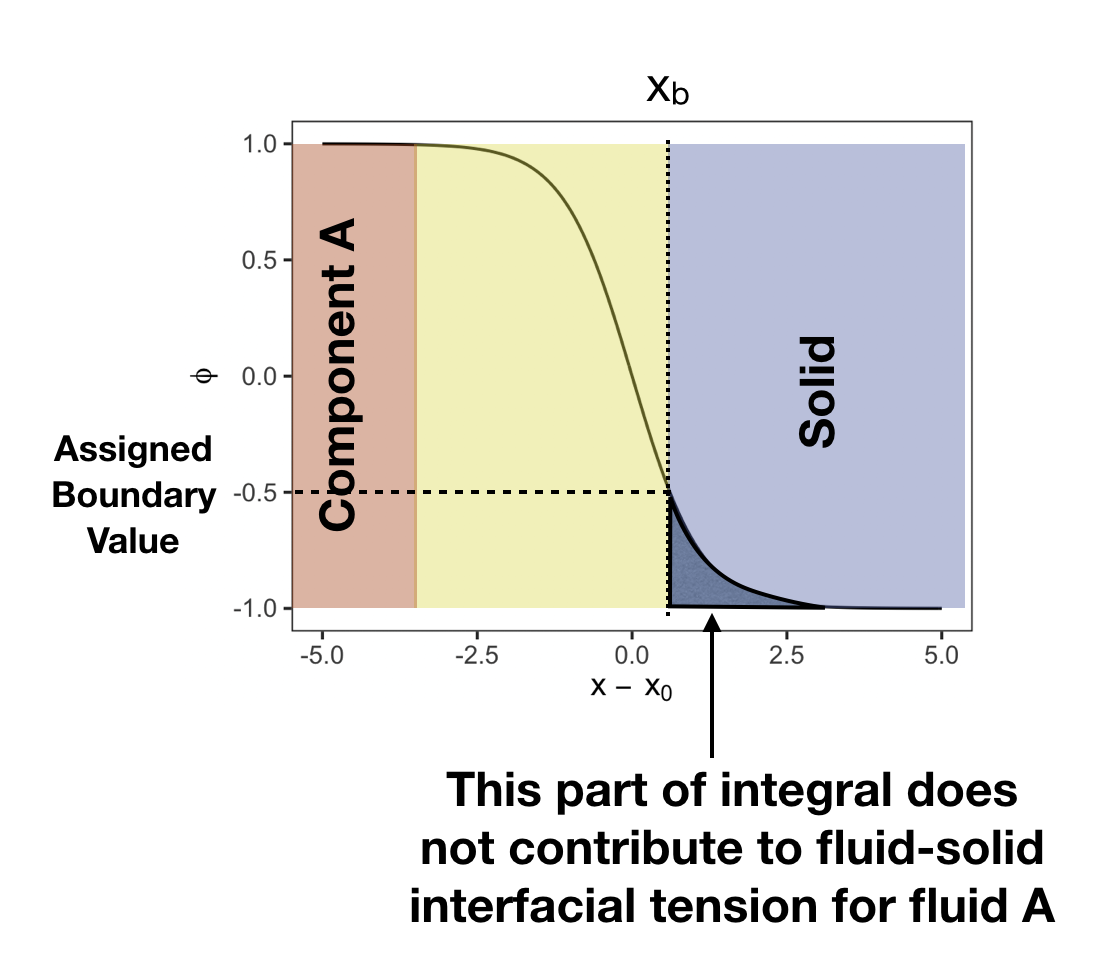}
\caption{Diffuse interface and wetting in color lattice Boltzmann model.
The fluid-solid interfacial tension is obtained by integrating the stress
profile across the diffuse interface. Assigning a constant boundary
value for the phase indicator field  
is an effective way to control the fluid-solid surface energy.}
\label{fig:diffuse-wetting}       
\end{figure}

The link between $\phi_s$ and the definition of the interfacial tension can be understood based on the mechanical definition established from molecular arguments. As in physical systems, the interface region in LBMs is diffuse. At the scale of the diffuse interface the interfacial tension is undefined, since it accounts for tangential stresses of the two-dimensional interface
that is constructed as an approximation based on the Gibb's dividing surface \cite{Adamson_Gast_97}. The interfacial tension can therefore be defined
by integrating the associated stresses over the entire thickness of the interface.
In the color LBM the equilibrium interface profile can be described by
\cite{McClure_Prins_etal_14}:
\begin{equation}
    \Phi_a(\bm{x}) = 
    \frac{1}{2\beta} \log{\Bigg( \frac{1 + \phi(\bm{x})}{1 - \phi(\bm{x})} \Bigg)}.
    \label{eq:lb-dist}
\end{equation}
where $\Phi_a(\bm{x})$ is the distance to the Gibb's dividing surface, 
defined here as $\phi(\bm{x})=0$. The convention is that $\Phi_a$ is
positive within the part of space occupied by fluid $a$.
The equilibrium interface shape is depicted in 
Fig. \ref{fig:diffuse-wetting} for the situation where fluid component $a$
is in contact with the solid. For illustrative purposes 
the phase indicator field is assigned as $\phi_s = -0.5$ within the solid region.
The fluid-solid interfacial tensions can thus be computed from the mechanical
definition,
\begin{eqnarray}
\sigma_{as} &=& \int_{-\infty}^{x_b} (\sigma_{xx} -\sigma_{yy}) dx \;, \\
\sigma_{bs} &=& \int_{x_b}^{\infty} (\sigma_{xx} -\sigma_{yy}) dx \;,
\end{eqnarray}
where $\sigma_{xx}$ is the normal component of the stress tensor and
$\sigma_{yy}$ is the tangential component. Away from equilibrium, both
the interface profile and stress tensor can deviate from their equilibrium 
state, which is consistent with the intended behavior for a mesoscopic method.
To the extent that the underlying interaction rules used to construct the method
have a reasonable physical basis, the non-equilibrium deviations are also physically
reasonable. As an example, the color LBM naturally reproduces appropriate 
scaling behavior for the moving contact line, i.e. recovery of the well-known
Tanner-Cox-Voinov rule for the dynamic contact angle \cite{Latva_Kokko_Rothman_07}.

For strongly immiscible fluids, the interface thickness will typically be $1.0$--$5.0$ nm. In LBMs, the thickness of the interface is determined from the image resolution. Since $\mu$CT image resolution is usually between $0.5$--$20$ $\mu$m, the diffuse interfaces in a typical digital rock simulation will be several orders of magnitude thicker than their physical counterparts. This can be problematic, particularly if the interface thickness approaches the size of a typical pore throat in the system \cite{doi:10.1002/fld.4822}. On the other hand, reasonable behavior should be expected provided that the dominant flow channels are sufficiently well-resolved. After all, the color LBM provides an adequate basis to model geometric (i.e. contact angle) effects for both static and dynamic cases, which account for essential first-order effects. The basis to model composition effects within the interface region, such as mass transfer within films, is less well established. Alternative mesoscopic models, such as those based on Cahn-Hilliard theory or free energy, do provide a basis to study the interface composition \cite{Seppecher_96,Lee_Lin_08,Zhang_Kwok_09}. However, since the associated effects are less important as compared to geometric effects and associated connected pathway flow, we do not consider these details in the present study. The basis to exclude these effects is that the dominant flow channels are sufficiently well-resolved in comparison to the thickness of the diffuse interface.

\subsection{Initial conditions and wetting maps}
\label{sec:initial}

The input geometry for simulation is provided based on a labeled image where each voxel is assigned a particular integer value to denote the occupying material component. Typical segmented $\mu$CT can be directly ingested by the simulator
to assign the position of both fluid and solid components. In most cases, $\mu$CT 
data does not include the spatial distribution of fluids, which then must be assigned based on some other mechanism. LBPM includes capabilities to initialize fluid configurations based on morphological analysis of the pore structure. 

\begin{figure}
\includegraphics[width=0.7\textwidth]{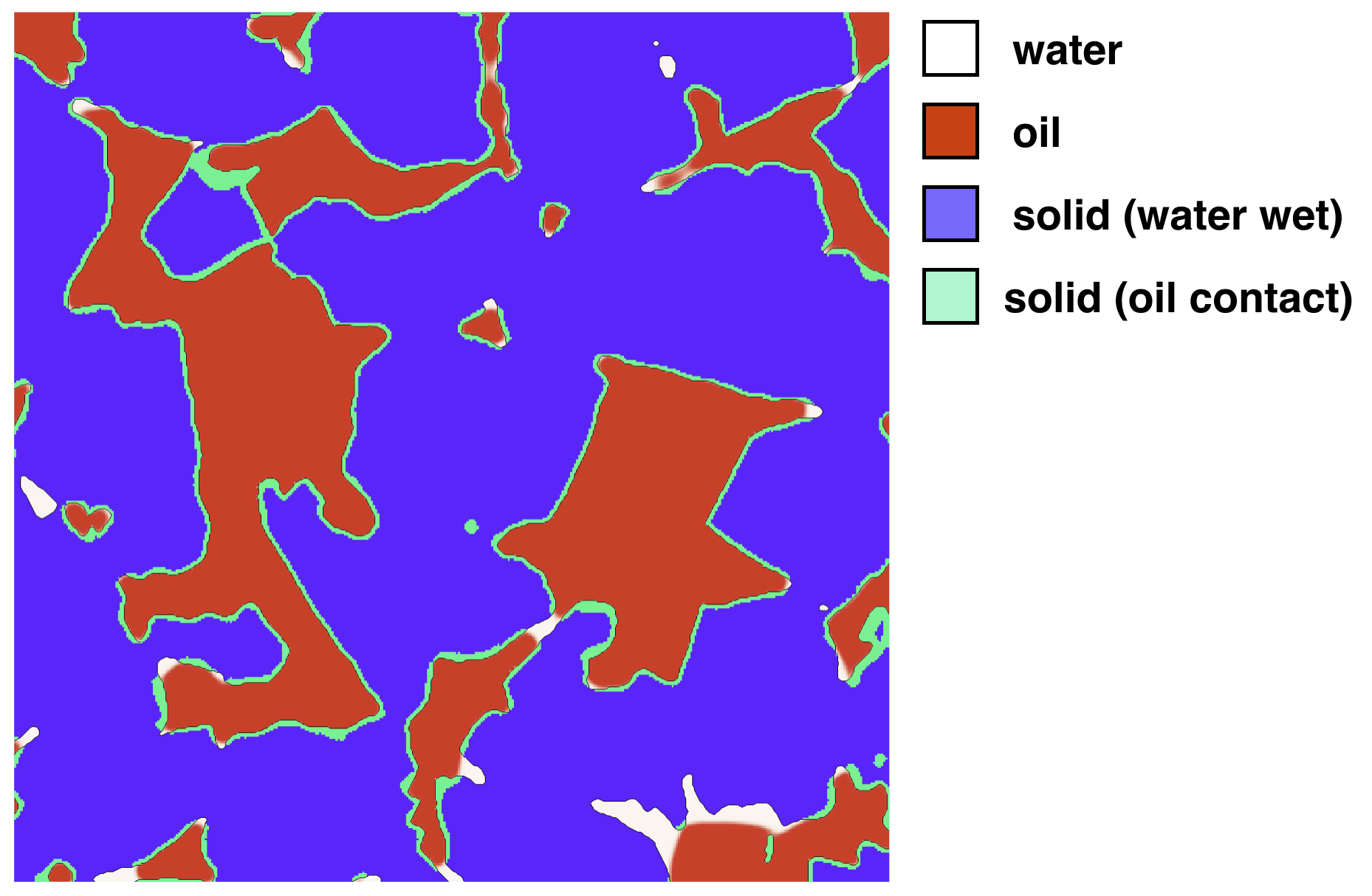}
\caption{Initial conditions for two-fluid flow simulation are initialized by  applying the morphological drainage operation within the pore space to obtain an initial configuration for the non-wetting fluid (red) and water (white) within the solid matrix; portions of the solid that contact oil are assigned a new label so that mixed-wet behavior can be modeled. 
}  
\label{fig:wetting-map}
\end{figure}

Fluid configurations along primary drainage can be efficiently generated
based on the morphological drainage operation \cite{Hilpert_Adalsteinsson_06}. Although this approach is particularly effective for modeling drainage under strongly water-wet conditions, it remains a useful tool within the context of modeling more complex wetting behaviors. A good example are ``mixed-wet" systems, where part of the solid surface is water-wet and part is oil-wet, with the spatial distribution being assigned based on the fluid history. Conceptually, the rocks begin as water-wet, with only water present in the system.
Oil is added based on a drainage process, which displaces water such that it remains only in the corners. Over time, the surface wetting properties are altered for those parts of the solid surface that contact oil. The associated surface energy becomes more favorable toward oil based on an aging process.
To realize a mixed-wet condition and to mimic an aging process, we utilize the following procedure:
\begin{enumerate}
    \item initialize the configuration of fluids based on a primary drainage operation with a desired water saturation;
    \item compute a distance transform relative to the morphologically instantiated oil phase;
    \item re-label all solid voxels within $2$ voxel units of the oil with a new label.
\end{enumerate}
An example configuration is shown in Fig. \ref{fig:wetting-map}. The re-labeled image
can then be used to assign the interfacial energy based on the desired spatial distribution of wettability.

\subsection{Analysis framework}
\label{sec:analysis}

Simulation of multiphase flows routinely requires a large number of timesteps,
and analyzing the associated simulation results can itself presents a challenging computational task. LBPM includes a thread-based framework to carry out {\em in situ} analysis of the flow behavior \cite{McClure_Wang_etal_2014}.
The analysis framework is configured to compute integral measures from
the flow that capture essential aspects of the behavior. The inputs
for the analysis are the pore-scale fields obtained from the lattice Boltzmann simulator. The phase indicator field $\phi(\mathbf{x})$, identifies the location of the two fluids within the rock (i.e. $\phi(\mathbf{x}) > 0$ for the region occupied by water and $\phi(\mathbf{x}) < 0$ for the remaining part of the pore space filled by oil), as well as an interface region also identified as the part of the system where significant gradients in composition are observed, $|\nabla \phi| \ge \epsilon$, where $\epsilon=0.001$. Each of these regions can be further sub-divided into connected and disconnected parts based on connected components analysis \cite{McClure:2016:ASC:3018859.3018862}. The resulting sub-regions are listed in Table \ref{tab:entities}.
Each sub-region corresponds to a geometric entity that can be used 
as the domain for integration when computing integral measures. 
The boundary for each sub-region is constructed explicitly as a list
of triangles based on a double connected edge list (DCEL) data structure. 
The DCEL data structure can be conveniently exploited to compute
invariant quantities such as the Minkowski functionals
\cite{Schr_der_Turk_2013}. Within LBPM the scalar geometric invariants
are computed for each sub-region $\alpha \in \{wc,wd,nc,nd,ic,id\}$:
volume $V_\alpha$, surface area $A_\alpha$, integral mean curvature
$H_\alpha$ and Euler characteristic $\chi_\alpha$.

\begin{table}
\caption{Geometric entities constructed within the LBPM analysis framework.}
\label{tab:entities} \begin{tabular}{lll}
\hline\noalign{\smallskip}
Symbol & Boundary & Description of entity  \\
\noalign{\smallskip}\hline\noalign{\smallskip}
$\Omega_{wc}$ & $\Gamma_{wc}$ & connected part of the water \\
$\Omega_{wd}$ &$\Gamma_{wd}$ & disconnected part of the water \\
$\Omega_{nc}$ & $\Gamma_{nc}$ & connected part of the oil \\
$\Omega_{nd}$ & $\Gamma_{nd}$ & disconnected part of the oil \\
$\Omega_{ic}$ & $\Gamma_{ic}$ & connected part of the interface region \\
$\Omega_{id}$ & $\Gamma_{id}$ & disconnected part of the interface region \\
\noalign{\smallskip}\hline
\end{tabular}
\end{table}

For each region, averages of physical quantities can be determined 
based on the fluid density $\rho(\mathbf{x})$, pressure $p(\mathbf{x})$ and velocity $\mathbf{u}(\mathbf{x})$. 
The associated integral measures are listed in Table \ref{tab:integral-measures}, which include the total mass, momentum and kinetic energy for each region. It is mindful to 
include the total amounts, since these quantities must obey associated global conservation laws.
Averaged quantities can easily be determined from the totals. For example, the average flow 
velocity for fluid $\alpha$ is determined as:
\begin{equation}
    \big< \mathbf{u} \big>_\alpha = \frac{\mathbf{P}_\alpha}{M_\alpha}.
\end{equation}
Similarly we can compute the average flow rate for the two-fluid system 
by summing the total mass and momentum for both fluids:
\begin{equation}
    \big< \mathbf{u} \big> = \frac{\mathbf{P}_b + \mathbf{P}_a}{M_b + M_a}.
\end{equation}
Comparing the previous two equations it is clear that, to be consistent with momentum balance, the total flow rate cannot be defined as the sum of the flow rate for both fluids. It is mindful to directly report the totals for conserved quantities, since this allows for the exploration of different kinds of averages more easily. The average pressure is also defined for each fluid. Since the product of the pressure and volume contributes to the internal energy of the system, the average pressure is computed such that $p_\alpha V_\alpha$ provides the associated contribution in accordance with classical thermodynamics. 

\begin{table}
\caption{Integral measures computed by the LBPM analysis framework.}
\begin{tabular}{lll}
\hline\noalign{\smallskip}
Symbol & Definition & Description of entity  \\
\noalign{\smallskip}\hline\noalign{\smallskip}
$V_\alpha$ & $\int_{\Omega_{\alpha}} dV$ & total volume for region $\alpha$ \\
$A_\alpha$ & $\int_{\Gamma{\alpha}} dS$ & total surface area for region $\alpha$ \\
$H_\alpha$ & $\int_{\Gamma{\alpha}} (\kappa_1 +\kappa_2) dS$ & integral mean curvature for region $\alpha$ \\
$\chi_\alpha$ & $\mathcal{V}_\alpha - \mathcal{E}_\alpha + \mathcal{F}_\alpha $& Euler characteristic for region $\alpha$ \\
$M_\alpha$ & $\int_{\Omega_{\alpha}} \rho dV$ & total mass for region $\alpha$ \\
$\mathbf{P}_\alpha$ & $\int_{\Omega_{\alpha}} \rho \mathbf{u} dV$ & total momentum for region $\alpha$ \\
$K_\alpha$ & $\int_{\Omega_{\alpha}} \rho \mathbf{u} \cdot \mathbf{u}  dV$ & total kinetic energy for region $\alpha$ \\
$p_\alpha$ & $\frac{1}{V_\alpha}  \int_{\Omega_{\alpha}} p dV$ & average pressure for region $\alpha$ \\
\noalign{\smallskip}\hline
\label{tab:integral-measures}
\end{tabular}
\end{table}

Combining the entities summarized in Table \ref{tab:entities} and with the measures in Table \ref{tab:integral-measures}, provides the basis to analyze a wide range of subphase behaviors, from both the structural and physical
standpoint. The associated measures are logged to a space-delimited CSV file {\tt subphase.csv} at a user-specified
time-step interval. The subphase analysis capabilities provides a convenient way to track how flow behavior evolves with time in large complex digital rock images. Key opportunities are to explore the role of connected pathway flow,
transient changes in fluid topology, ganglion dynamics, and the role of film structure in transport. 

\section{Core analysis workflows}
\label{sec:SCAL}

In addition to the subphase analysis described above, LBPM also includes an analysis protocol that can be used to track a smaller set of basic quantities including fluid saturation, effective permeability, average velocity and average pressure for each fluid.
The basic analysis avoids more expensive computational routines such as interface construction and connected components analysis such that more frequent analysis can be performed without sacrificing simulation performance.
LBPM relies on these basic analysis routines to detect numerical instabilities, which allows for
unstable simulations to be terminated efficiently. Both analysis protocols can be enabled simultaneously to provide different temporal granularity. For simulations performed in this work, the basic analysis was performed every
$5000$ timesteps to provide near real-time feedback on the progress of each simulation. 

\subsection{Simulation protocol for unsteady flow}
\label{sec:unsteady-protocol}

\begin{figure}
\includegraphics[width=1.0\textwidth]{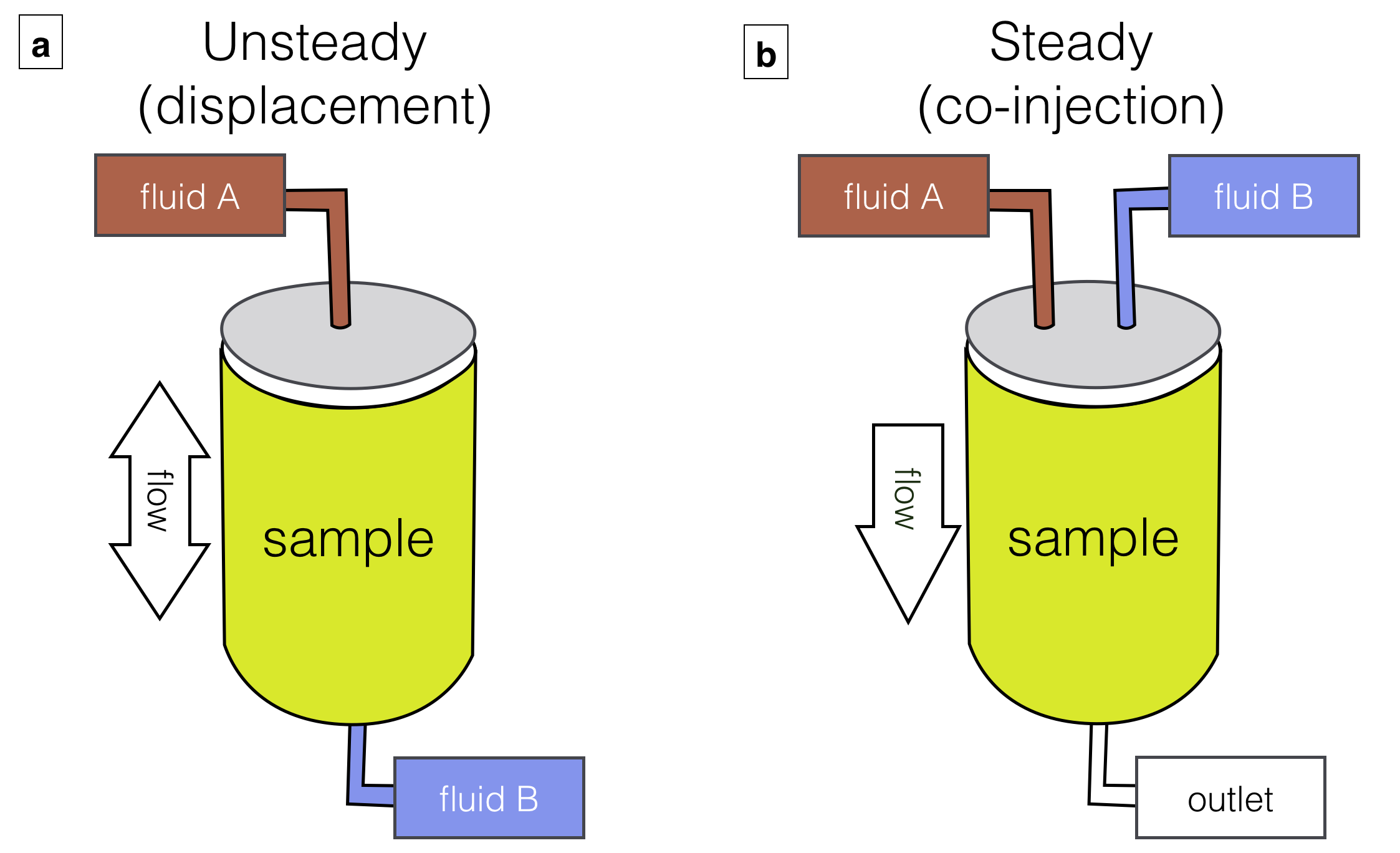}
\caption{ Illustrative examples of common approaches for special core analysis laboroatory (SCAL) experiments:
(a) fluid reservoirs are established at opposite ends of the sample core;
with one fluid displacing the other based on the experimental conditions;
(b) two fluids are co-injected at the inlet to obtain a steady-state fractional flow rate.
}  
\label{fig:SCAL_workflow}
\end{figure}

The traditional experimental approaches to measure flow behavior within rock have been
refined into a variety of special core analysis laboratory (SCAL) workflows.
In special core analysis, flooding experiments are routinely performed
to study the behavior of the capillary pressure and assess the endpoints.
A common situation is to displace one fluid with another, establishing a reservoir of one fluid at the inlet and another fluid at the outlet, as illustrated in Fig. \ref{fig:SCAL_workflow}a.
Either fluid can be injected to displace the other depending on the experimental conditions. The associated experiments are typically unsteady because the volume fraction of each fluid changes as fluid
is injected. LBPM can be used to run analogous simulations by applying 
external boundary conditions to drive the flow behavior. Two boundary
conditions in particular are useful for designing such computational
experiments. First is the basic pressure boundary condition, which 
can be used to assign constant pressure values to the inlet and
outlet \cite{Ginzburg_etal_08,Zou_He_1997}. Alternatively, a flux
or velocity boundary condition can be applied to assign the flow 
rate \cite{Li_McClure_fluxb_BC}. The flux boundary condition is distinct
from conventional velocity boundary conditions because it sets a constant
total volumetric flow rate across the entire inlet instead of enforcing a constant velocity at each point at the inlet. This condition improves numerical stability by allowing the boundary velocity to adapt dynamically to the interior flow, ensuring that the applied boundary condition is consistent with the behavior of the simulated system; it is similar to many physical experiments where the total flow rate is known based on the rate at which fluids are injected into the system.

Computational experiments can be also be used to construct a 
workflow that is analogous to centrifuge experiments used to measure capillary pressure curves. 
In centrifuge experiments the sample is attached to an
arm that rotates about an axis, imposing a centrifugal force on the system based on the length of the axis arm,
rotation rate, and the density difference between the two fluids. The more dense fluid will tend 
to be driven toward the outlet. The distribution of fluids remaining within the sample is determined
based on the balance of capillary forces against the centrifugal forces, providing a useful 
mechanism to infer the capillary pressure. Depending on the surface wetting properties of the sample, the situation shown in Fig. \ref{fig:SCAL_workflow}a will typically lead to spontaneous imbibition of one fluid or the other if pressure conditions are used to enforce zero pressure drop across the sample. If an external body
force is also applied, the balance of forces in the system will be altered and fluid will be mobilized.
This is very similar to a centrifuge experiment described above. Once the fluid saturation stabilizes, the external body force can be used to approximate the capillary pressure as measured in a centrifuge.
Unlike a centrifuge experiment, the computational approach will work well even for fluids with identical
mass density.

\subsection{Simulation protocol for steady-state flow}
\label{sec:steady-state-protocol}

The steady-state simulation protocols in LBPM combine two basic components
that are needed to efficiently generate ``steady-state" multiphase flows in
porous media:
\begin{enumerate}
    \item a mechanism to drive and monitor the flow behavior; and
    \item a mechanism to modify the fluid configuration and fluid saturation in particular. 
\end{enumerate}
To study multiphase flows, an apparatus must be constructed to inject fluids into the sample so that constitutive relationships can be explored for different situations. A common experimental set-up is illustrated in Fig. \ref{fig:SCAL_workflow}b. To induce a quasi-steady state flow
as needed to measure relative permeability, two fluids are co-injected at the inlet (e.g. using syringe pumps) and flow through the sample at a desired fractional flow rate. By monitoring the pressure drop across the sample, the relative permeability can be calculated from the experimental measurements.
In a typical experimental setup the flow behavior is controlled by manipulating the fractional flow behavior\cite{avraam_payatakes_1995}. In many situations
such experiments may not generate truly steady-state flows due to naturally-arising fluctuations in the fluid saturation. In this respect computational algorithms have an apparent advantage, since flow at constant saturation can be directly imposed by using 
a periodic boundary condition while driving the flow with an external driving force, a common approach in digital rock based simulation of
relative permeability.

\begin{figure}
\includegraphics[width=1.0\textwidth]{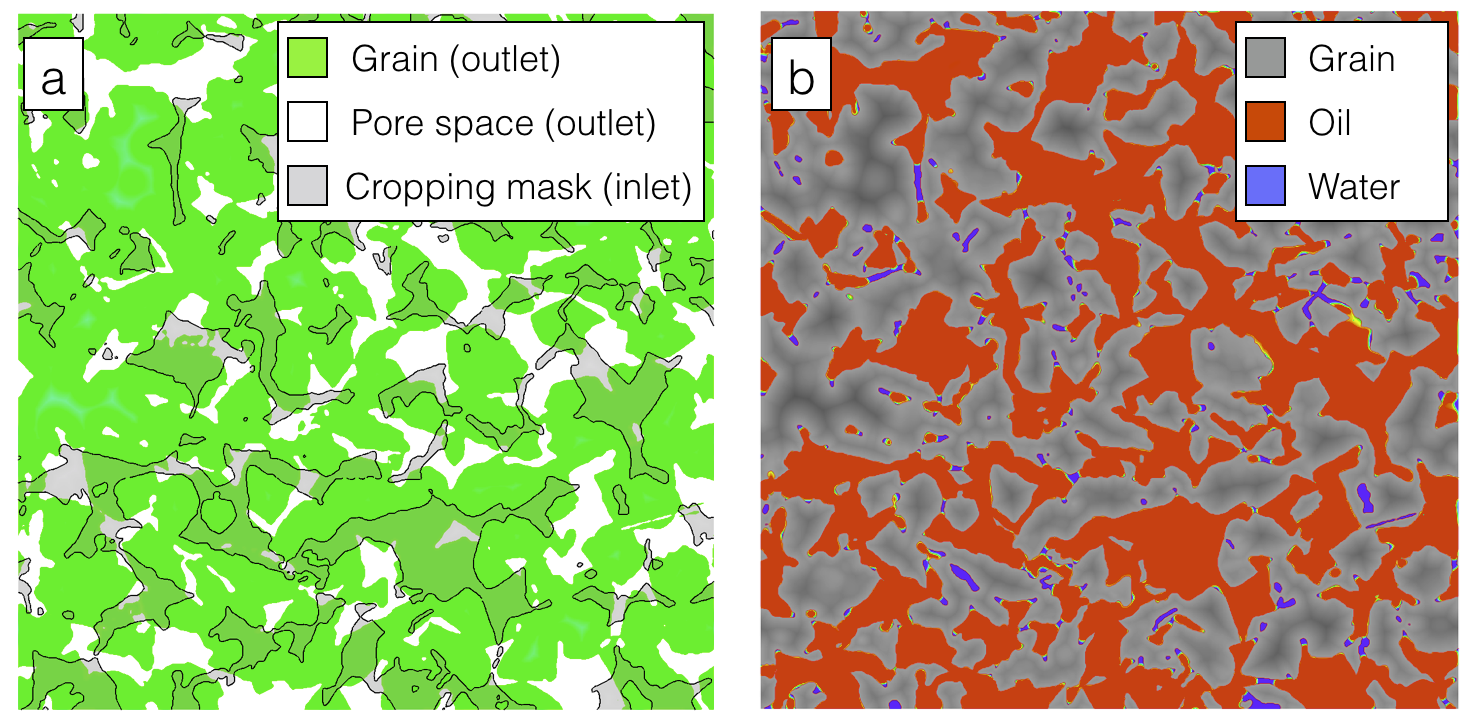}
\caption{ Construction of outlet mixing layer within Bentheimer sandstone:
(a) a cropping mask is generated based on the shape of the porespace
at the inlet (transparent grey); at the outlet any solid grain (green) that intersects with cropping mask is removed;
(b) the resulting mixing layer prevents the mismatch at the inlet / outlet
from blocking flow pathways, supporting both large flow conduits through the large pores and film flow conduits along the grain structure. An
analogous condition is applied at the inlet.
}  
\label{fig:mixing}
\end{figure}

Imposing a periodic boundary condition in a real digital rock image presents a
challenge due to the fact that the underlying rock structure is not periodic. 
This can lead to serious problems because the boundary mismatch can 
block flow pathways and artificially reduce the effective permeability. Various
strategies can be considered to address this, such as:
(1) double the domain length by adding a mirror-image as a way to enforce periodicity (e.g. \cite{ISI:000450094200004});
(2) implement boundary conditions that directly mimic the symmetry of 
domain reflection (e.g. \cite{Fredrich_Digiovanni_etal_06}); and
(3) create a mixing region at the boundary so that the two fluids can establish a pathway through the system.
In this work we consider the third option, where the mixing region is
within $N_{mix}$ voxels of the boundary. An overview of the approach
used to construct the mixing is provided in Fig. \ref{fig:mixing}.
Within the mixing region, the pore structure at the inlet and outlet 
are used as a mask to 
remove solid voxels so that they do not block the flow. Along the inlet,
the mask is obtained based on the pore structure at the outlet face; all solid voxels that overlap with the porespace region at the outlet are
removed. The opposite convention is applied at the outlet, removing 
any solid voxels that block the inlet pore structure. The resulting
mixing region guarantees that flow pathways will not be blocked artificially
(at the expense of artificially increasing the porosity). Any solid wetting labels present
in the original image can be preserved based on the transformation, 
and the connectivity of film structures can also be maintained across
the boundary. In the results section we show that this rule can reduce
the impact of boundary effects.

The capillary number characterizes the dominant physical regime for pore-scale multiphase flows based on the balance of capillary and viscous forces. The conventional definition is ambiguous when two fluids are present, since both the viscosity and flow rate can vary independently. Within LBPM we define a total capillary number based on the sum of the capillary numbers obtained for each fluid:
\begin{equation}\label{eq:capillary_number}
    \mbox{Ca} =  \frac{\epsilon_a \mu_a |\bm{u_a}| + \epsilon_b \mu_b |\bm{u_b}|}{\sigma}.
\end{equation}
where $\epsilon_a$ and $\epsilon_b$ are the volume fraction of the fluid components
and $\bm{u_a}$ and $\bm{u_b}$ are the average rate of momentum flux within each fluid region based on the definitions given in \S \ref{sec:analysis}. The total capillary number defined in Eq. \ref{eq:capillary_number} is advantageous because it retains a representative value even if the
flow rate within one of the fluids is effectively zero, as is the case when the connectivity of one fluid breaks down.  

Steady-state flow is driven by an external body force specified by the user.
The body force can optionally be adjusted dynamically to match a target capillary 
number $\mbox{Ca}^*$. If a target capillary number is specified, LBPM will internally
rescale the body force based on the measured flow rate:
\begin{equation}\label{eq:LB_body_force_adaptation}
    \bm F \leftarrow \frac{\mbox{Ca}^*}{\mbox{Ca}} \bm F.
\end{equation}
The correction to the capillary number is designed to insulate the forcing term from the undue impact of noise within the system. To effectively accomplish this goal,
the capillary number must be measured based on a flow that is somewhat representative
of the final steady-state. For a particular configuration such a flow can be generated 
by allowing the steady-state flow to relax for a user-specified number of timesteps
before the body force is rescaled.  In this work a timestep interval of $100,000$
timesteps was applied. The force is also rescaled after each steady-state point is
obtained for situations where multiple points are obtained in sequence. Since the
response of the flow rate to the force can be non-linear as the capillary 
number increases, the algorithm is implemented to cap the rescaling so the force
is never changed by more than a factor of two. This is necessary to 
prevent undesirable oscillations that result as a consequence of
non-linearities in the flow behavior. 

Once the external driving force has been determined, steady-state relative permeability calculations are performed. Steady-state is assessed
by considering the change in the capillary number based on a user-defined
timestep interval:
\begin{equation}
    \Delta\mbox{Ca}  = \left| \frac{\mbox{Ca}(t)-\mbox{Ca}(t-t')}{\mbox{Ca}(t)} \right|,
\end{equation}
where the time interval $t' = 100,000$ timesteps for results presented
in this work. Steady state is achieved when $\Delta\mbox{Ca}$ is less than a user specified tolerance. Due to various non-linear effects, the approach to steady state is frequently non-monotonic for multiphase flows. 

\begin{figure}
\includegraphics[width=1.0\textwidth]{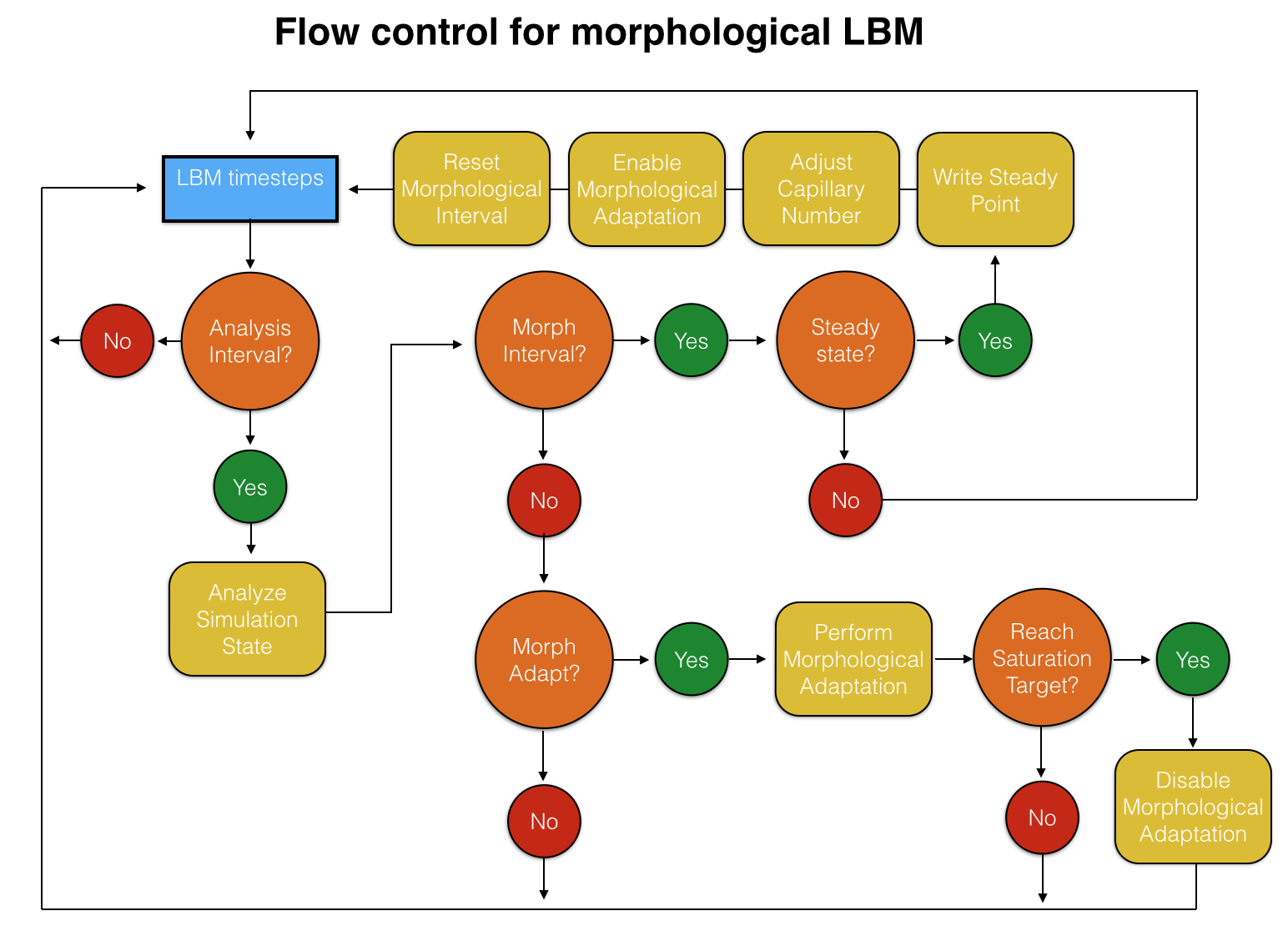}
\caption{ LBPM flow control diagram governing the assessment of steady state and application of the morphological routine in conjunction with the two-phase lattice Boltzmann approach.
}  
\label{fig:flow}
\end{figure}

LBPM provides several mechanisms to vary the fluid configuration, 
which are presented as internally-supported simulation protocols. The most flexible of these is an {\it image sequence} protocol that allows the user to provide a list of 3D images to specify the fluid configuration to be used as the input for the steady-state algorithm. The provided images may be segmented
$\mu$CT images of real fluid configurations \cite{PhysRevE.94.043113}, outputs of previous steady or unsteady
LBPM simulations, morphologically generated fluid configurations, or any 
other user-generated volumetric data that specifies the fluid distributions. 
Additional protocols are available within LBPM for which the fluid configurations are modified internally based on customized algorithms. The {\em shell aggregation} protocol is 
described in detail in the following sections. Fig. \ref{fig:flow} 
depicts the flow control procedure within LBPM. The simulation state
is analyzed at regular intervals, which is used to inform routines that
perform morphological adaptation and adjust flow parameters.

\subsubsection{Image sequence protocol}

As input, LBPM requires a labeled 3D image that segments the computational domain
based on the material type. The {\em image sequence} protocol allows for sequence of 
input images to be provided and used as the basis for a sequence of steady-state simulations subject to the constraint that the solid micro-structure specified within each image must be the same. A number of previously published works rely on an analogous workflow to simulate steady-state two-fluid flow behavior in digital rock images: multiple authors have explored the direct initialization of steady-state flow simulations from images generated from simulations of unsteady displacement \cite{ISI:000307391900003,ISI:000450094200004,ISI:000277220600014};
Fan et al. initialized fluid distributions using morphological tools 
\cite{FAN2019522};
Armstrong et al. initialized steady-state simulations from fast $\mu$CT images of fluid distributions observed from fractional flow experiments \cite{PhysRevE.94.043113}. 
It is noted that there are many possible ways to instantiate fluids into a porous system,
and that many valuable studies can be performed by ingesting a sequence of images and
running simulations. The {\em image sequence} protocol generalizes and streamlines the procedure for conducting such studies so that the number of manual steps can be reduced.

\subsubsection{Shell Aggregation Protocol}

The only way that the volume of object can change is based on movement of the boundary.
The {\em shell aggregation} protocol relies on this basic fact to modify the arrangement of fluids using small boundary movements; then LB simulation is applied to the resulting fluid configurations to cause them to relax toward minimal energy configurations \cite{WANG2020108966}. Morphological methods have been used for the study 
of porous media for nearly two decades as a way to approximate the geometric structure of fluids  \cite{Pan_Hilpert_Miller_PRE_01}. A key advantage of the morphological approach is that it is much less computationally intensive as compared to direct numerical simulation. On their own,
morphological methods perform relatively well to approximate the fluid geometry for conditions that are both capillary-dominated and strongly water-wet. Their utility is limited when viscous forces are important or under mixed-wet conditions. In such cases LBMs are needed to deliver reliable modeling results. To utilize both the computational efficiency of morphological approaches and the robust physics given by lattice-Boltzmann methods, here we rely on a highly-efficient hybrid approach: the direct LB simulation is embedded with a morphological adaptation which induces small changes in the fluid volume based on incremental movement of the boundary position, after which LB simulation is applied to adjust the phase configuration according to the particular physics. 

We develop a ``morphological growth" algorithm that is used to modify the arrangement of fluids and re-initialize the simulation at a different fluid saturation: 
\begin{enumerate}
    \item Copy the phase indicator field $\phi$ obtained from lattice Boltzmann simulation which indicates the position of fluids;
    \item Generate a distance map $\Phi_a$ according to the position of fluid $a$;
    \item Change the volume of the object by applying a local morphological operation to dilate or erode the fluid-fluid interface, obtaining a new
     distance map $\Phi_a^\prime$ ;
    \item Re-initialize the phase indicator field and restart the LB simulation.
\end{enumerate}
For a given fluid component $a$, e.g. $a=\{n,w\}$ for a two-fluid system, we can also define an indicator function for fluid $a$ as:
\begin{equation}
\Upsilon_{a} (\bm{x}) = \left \{ 
\begin{array}{cc}
  1,   &  \mbox{ if $\bm{x} \in \Omega_{a}$ }\\
  0,   &  \mbox{ otherwise}
\end{array}
\right.
\end{equation}
The distance transform $\Phi_{a}$ can then be calculated to determine the minimum distance to the boundary of $\Omega_{a}$, so that $\Phi_{a}(\bm{x}) < 0$ inside the fluid, and $\Phi_{a}(\bm{x}) > 0$ outside the fluid. For the domain $\Omega_{a}$ occupied by fluid $a$, the morphological operation can be considered as rolling a small ball with diameter $\delta$ around the exterior boundary of $\Omega_{a}$ and add that part of space covered by the ball into $\Omega_{a}$. Denoting the ball as $\zeta_\delta$, we can refer the morphological dilation as $\Omega_{a} \oplus \zeta_\delta$. Likewise, we can consider rolling the ball around the interior of the boundary and remove this part from $\Omega_{a}$, which is the morphological erosion, denoted by $\Omega_{a} \ominus \zeta_\delta$. These two operations can be efficiently implemented based on the distance transform $\Phi_{a}$ \cite{Hilpert_Adalsteinsson_06}. Given that distance transform $\Phi_{a}(\bm{x})$ is the signed distance to the boundary of $\Omega_{a}$ at location $\bm{x}$, it is easily seen that: 
\begin{eqnarray}
    \Omega_{a} \oplus \zeta_\delta &=& \{\bm{x}:\Phi_{a}(\bm{x})  < \delta \}\; 
    \mbox{and} \\
    \Omega_{a} \ominus \zeta_\delta &=& \{\bm{x}:\Phi_{a}(\bm{x})  < -\delta \}.
\end{eqnarray}
Based on this, it is useful to rely on the distance transform to facilitate morphological analyses, and the convention hereafter is that a positive $\delta$ corresponds to dilation and a negative $\delta$ corresponds to erosion. Within the implementation, 
the algorithm is developed based on the fractional growth in the volume of the structure,
\begin{equation}\label{eq:shell_aggregation_vol_change}
    \Delta = \frac{V(\Omega_{a} \oplus \zeta_\delta)}{V(\Omega_{a})}\;,
\end{equation}
where the growth factor $\delta$ is estimated by applying a dummy growth factor
of one voxel to the structure, then linearly rescaling $\delta$
to estimate the value needed to produce the target volume change $\Delta$.
The process is repeated until the measured change in volume fraction exceeds $\Delta$.
LB timesteps are applied in between successive applications to allow the shape of the fluids 
within the porespace to relax. Due to overlaps with 
the solid, the algorithm will tend to over-predict the volume change, meaning that successive applications
are usually necessary. However, the computational burden associated with this portion of the algorithm
is relatively small.

Conveniently, the distance to the interface can be expressed in closed form based on the density profile for the interface at equilibrium,
as given in Eq. \ref{eq:lb-dist}. This function is well-defined only in the interface region, i.e. $ -1 < \phi(\bm{x}) < 1$. The width of the interface region can be explicitly controlled by the LB color model parameter $\beta$. In this work $\beta = 0.95$ is used, 
which leads to an interface width of $\sim 5$ voxels. This means that Eq.\ref{eq:lb-dist} will provide a reasonable estimate of the distance within $2$--$3$ voxels of the interface. Combining this with an approach to determine the far-field distance provides $\Phi_{a}(\bm{x})$ everywhere in the domain $\Omega_{a}$. The advantage of Eq.\ref{eq:lb-dist} is that $\Phi_{a}(\bm{x})$ can be estimated with sub-voxel resolution in the vicinity of the interface. When the erosion operation is applied to the entire boundary of a fluid, the impact is to mimic a film-swelling effect in the vicinity of the solid. While it is possible to disable this behavior by setting a local rule, we find that allowing the film-swelling effect to occur is desirable for the modeling of systems with a wide range of wetting conditions.

The final step needed to update the LB phase field is to re-compute the phase indicator field based on the updated distance map. This is realized by substituting the morphologically altered distance transform $\Phi_{a}^\prime$ into Eq.\ref{eq:lb-dist} and solving for the updated phase indicator field:
\begin{equation}
    \phi^\prime (\bm{x}) = 
     2 \frac{e^{-2\beta \Phi_{a}^\prime(\bm{x})}}{1 + e^{-2\beta \Phi_{a}^\prime(\bm{x})}} - 1.
    \label{eq:lb-dist-to-phase}
\end{equation}
The fluid number densities $N_a$ and $N_b$ are then re-initialized to be consistent with $\phi^\prime (\bm{x})$. This also updates the amount of mass for each of the two fluids in way that is consistent with the desired morphological change. 

The overall flow control for the morphological algorithm is summarized in Fig. \ref{fig:flow}. Analysis of the simulation state is performed at a user-defined interval $t_{\mbox{analysis}}$ to compute averaged measures of the flow behavior \cite{McClure_Wang_etal_2014,McClure:2016:ASC:3018859.3018862}. The analysis interval is specified to provide detailed tracking for the system dynamics in the approach to steady state. A second user-defined interval $t_{\mbox{morph}}$ is specified to assess whether or not the system has reached steady state based on a convergence criterion. This interval should be an even multiple of the analysis interval. The convergence is assessed based on the change in the capillary number over the interval $t_{\mbox{morph}}$:
\begin{equation}
    \frac{|\mbox{Ca}(t) - \mbox{Ca}(t-t_{\mbox{morph}}) |}{\mbox{Ca}(t)} < \varepsilon_{\mbox{tol}},
    \label{eq:convergence}
\end{equation}
where $\varepsilon_{\mbox{tol}}$ is the desired tolerance. To prevent early termination,
$t_{\mbox{morph}}$ should allow for sufficient number of time steps as to allow for the changes that will occur in an unsteady flow. When steady state is achieved based on the criterion given by Eq.\ref{eq:convergence}, the algorithm resets to induce the next iteration of morphological changes. The switch to enable morphological adaptation is turned on after a steady-state data point is obtained. Morphological adaptation will remain enabled until one of two criteria are met: (1) the specified target saturation change is met; or (2) a user-specified maximum time steps elapse ($20,000$ time steps used in this work). The time-step ceiling is implemented to inhibit overly aggressive application of the morphological routine once the oil connectivity breaks down. 

\section{Results}
\label{sec:Results}

\begin{figure}
  \includegraphics[width=1.0\textwidth]{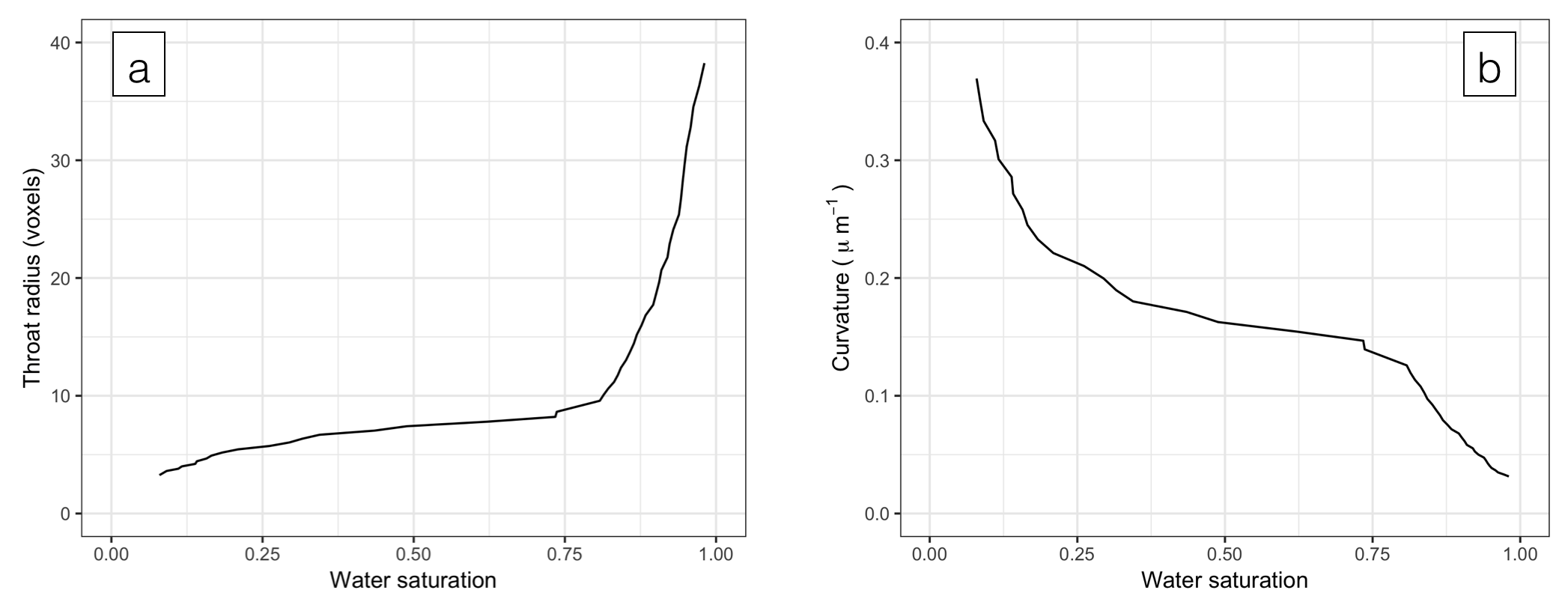}
\caption{Results obtained from morphological drainage in the Bentheimer sandstone:
(a) throat radius (in voxels) as a function of water saturation ; 
(b) morphological drainage curve analogous to mercury injection 
capillary pressure (MICP). 
}
\label{fig:morphdrain}       
\end{figure}

Simulations were performed using a $\mu$CT-imaged Bentheimer sandstone with a resolution 
of $1.66$ $\mu$m \cite{Dalton_2019}. A $900\times 900 \times 1600$ sub-region of the original image was extracted for flow simulations, which removed portions that
were outside of the cylindrical core-holder (refer to Fig. \ref{fig:uCT}). 
The porosity of the sub-region was $0.235$ and the absolute permeability was
$K = 1850$ mD. All simulations were performed with density and viscosity ratio
of one. In the simulation, the surface tension was set by setting $\alpha=0.005$.
Results were scaled to physical units based on an interfacial tension of 
$25$ mN/m, which is a typical value for oil-water systems. To assess
the image resolution, we applied the morphological drainage operation to classify the distribution of
pore throat sizes in the context of the image resolution. These results are presented in Fig. \ref{fig:morphdrain}. Key information about the image resolution is most 
easily inferred from Fig. \ref{fig:morphdrain}a, which shows how the pore throat radius 
changes with fluid saturation during morphological drainage. A smaller radius of curvature is obtained as the meniscus is pushed into smaller pore throats within the system.
The radius of curvature therefore corresponds to the minimum resolution for the dominant flow pathways at a particular saturation. In the Bentheimer sample, $R=5$ voxels at $S_w=0.15$, meaning that the diameter of the minimum occupied pore throat is $10$ voxels. With the throat radius significantly larger than the interface width, the meniscus location within the pore throats can be adequately resolved and the underlying physics can be simulated accurately. The resolution of this digital rock image is relatively high, which ensures that flow pathways are well-resolved across a wide range of saturation values.
Additionally, the morphological drainage operation establishes that the pore-space is well-connected, since the generated non-wetting phase corresponds to a single connected component. Morphological drainage provides a computationally efficient mechanism to identify digital rock images that are under-resolved, or where the porespace connectivity breaks down. This information can be used to ensure that expensive simulations are not performed on $\mu$CT data that lacks the resolution needed
to adequately describe the pore structure. In general, the morphological drainage operation provides an excellent mechanism to assess image quality based on resolution. It can also be used
to infer the capillary pressure behavior under water-wet conditions. The 
result is analogous to what would be obtained based on mercury injection
capillary pressure (MICP) curve. The associated capillary pressure curve is shown in Fig. \ref{fig:morphdrain}b. 

\subsection{Unsteady displacement}
\label{sec:unsteady-displacement}

\begin{figure}
  \includegraphics[width=1.0\textwidth]{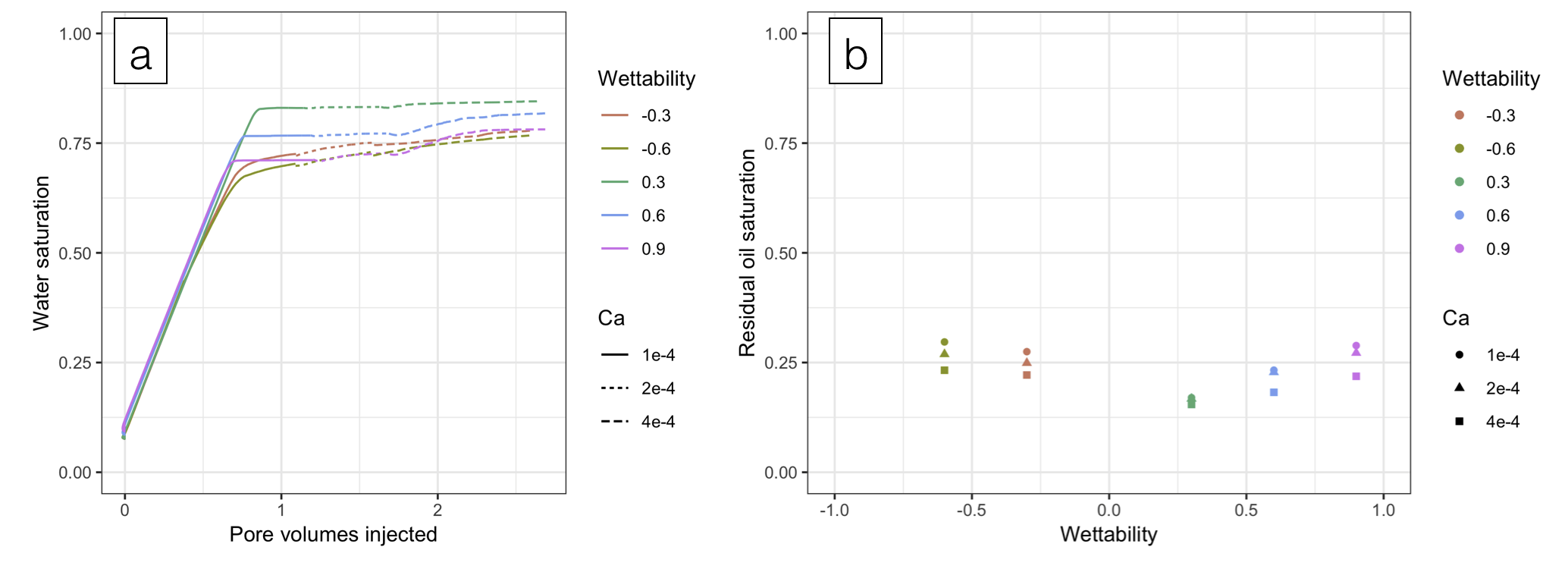}
\caption{Results of unsteady water flooding simulation in Bentheimer sandstone: (a)  water saturation as a function of injected water volume;
(b) residual oil saturation as a function of wettability at three different capillary numbers. 
}
\label{fig:waterflood}       
\end{figure}

For the unsteady water-flooding simulations, the initial condition for the simulation was established
based on the morphological drainage process with an initial water saturation $0.08$.
As described in \S \ref{sec:initial} all solid voxels initially in contact with 
oil were assigned a unique label in the input image. A mixed wet condition was
applied by assigning five different wetting conditions for the associated portion
of the solid surface, W$= \{0.9, 0.6, 0.3, -0.3, -0.6\}$. A strongly water-wet condition
was applied for the portion of the solid surface that did not initially contact with oil. Fig. \ref{fig:waterflood}a shows how water saturation evolves as water is injected into the system. The capillary number for the initial water injection was
$\mbox{Ca}=10^{-4}$. A constant slope is observed for the initial displacement
as an equal volume of oil is removed from the system to match the volume of injected
water. The slope changes once water establishes a connected pathway through the system,
which varies depending on the wetting condition. For the three water-wet cases, 
the water saturation is nearly constant after breakthrough. Additional oil is mobilized
only when the capillary number is increased based on two bump floods, each doubling
the capillary number to $2\times 10^{-4}$ and again to $4\times 10^{-4}$. For the two oil-wet cases, slight tail production is observed after breakthrough, which is consistent with experimental studies \cite{Jadhunandan_Morrow_1995}.

\begin{figure}
  \includegraphics[width=1.0\textwidth]{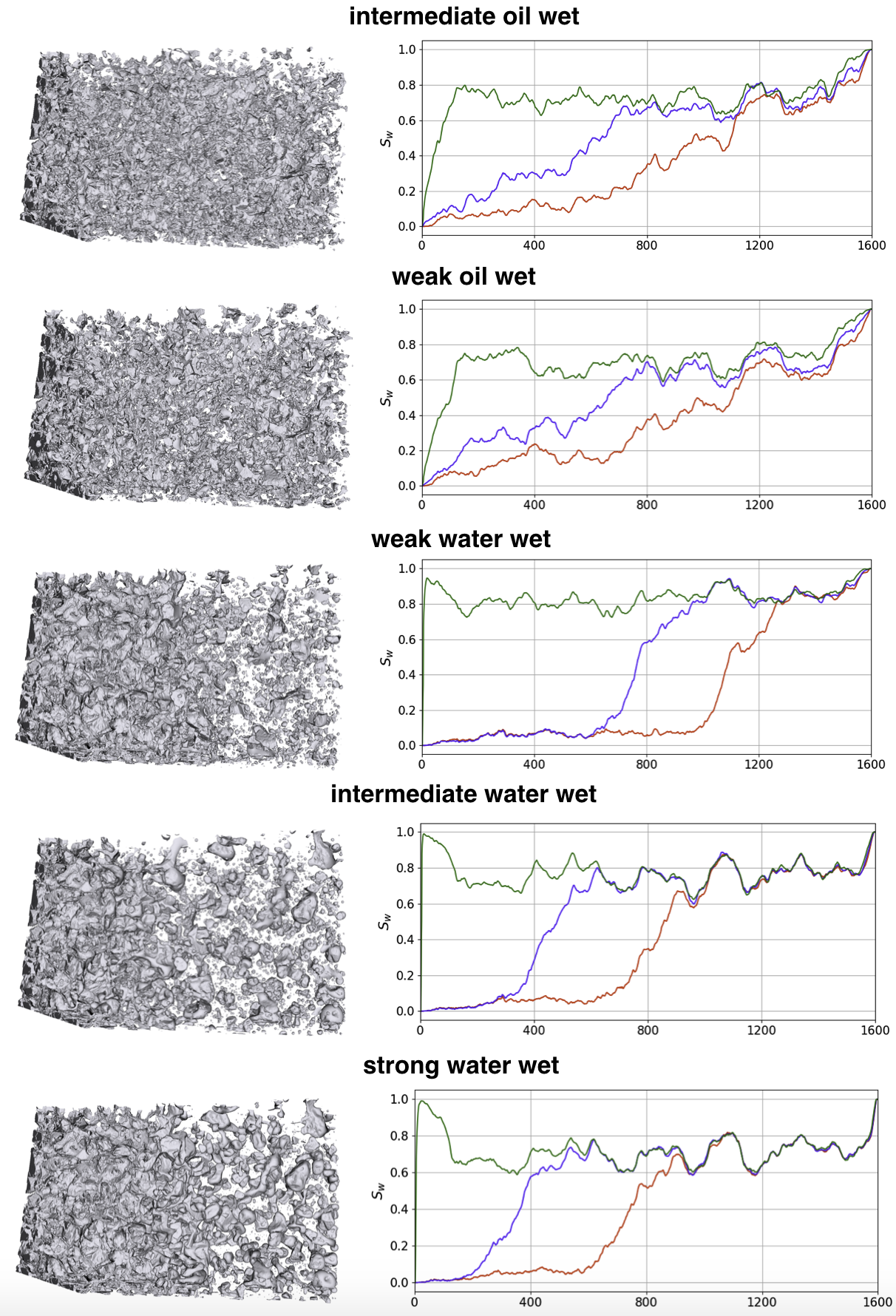}
\caption{ Simulations of water flooding using a flux boundary condition with $\mbox{Ca}=10^{-4}$: (left column) end-state oil distribution, where water and solid matrix are made transparent; (right column) local water saturation profiles along the flow axis, in which three representative global saturation of approximately 0.35, 0.45 and 0.72 is presented, corresponding to red, blue and green curves, respectively. It can be seen that local saturation profiles vary significantly based on position. A clear front is 
observed for the water-wet cases.}
\label{fig:saturation-profile-unsteady}       
\end{figure}

Local saturation profiles extracted from the unsteady displacement, as well as the associated end-state oil distributions, are shown in Fig. \ref{fig:saturation-profile-unsteady}. It is readily apparent that saturation
gradients are present during the displacement due to the fact that fluids must percolate
from one side of the domain to the other. A clear front is observed for the 
water-wet cases, which propagates across the system until the mobile oil is displaced. 
In these situations, the saturation is relatively constant both in front of and behind
the moving front. For more oil-wet cases, a drainage process is simulated and the associated
front is not sharp. In all cases, the saturation profile is quite homogeneous after breakthrough. The residual oil saturation can be inferred based on the saturation 
plateaus observed in Fig. \ref{fig:waterflood}a. Capillary number
dependence for the residual oil saturation is shown in Fig. \ref{fig:waterflood}b. 
A smaller amount of residual oil is observed for more neutrally-wet cases in the 
Bentheimer sandstone.

%

\subsection{Steady state relative permeability}
\label{sec:steady-state-relperm}

Simulations of steady-state relative permeability were performed based on the
{\em shell aggregation} protocol described in \S \ref{sec:steady-state-protocol}.
Initial conditions were generated based on the morphological drainage operation 
with water saturation $S_w=0.08$, as for \S \ref{sec:unsteady-displacement}. 
Following each steady-state point, shell aggregation was performed with 
a target volume change $\Delta  = -0.08$ (see Eq.\ref{eq:shell_aggregation_vol_change}) to mimic a water flooding process.
The analysis interval was $5,000$ timesteps and a maximum of $100,000$ 
timesteps spent within the morphological adaptation portion of the algorithm.
The target capillary number of simulations was $\mbox{Ca}=10^{-5}$, and the external
body force was re-scaled after $100,000$ timesteps for each steady-state point.
Steady-state relative permeability curves were generated for each of the five different mixed-wet conditions.

\begin{figure*}
  \includegraphics[width=1.0\textwidth]{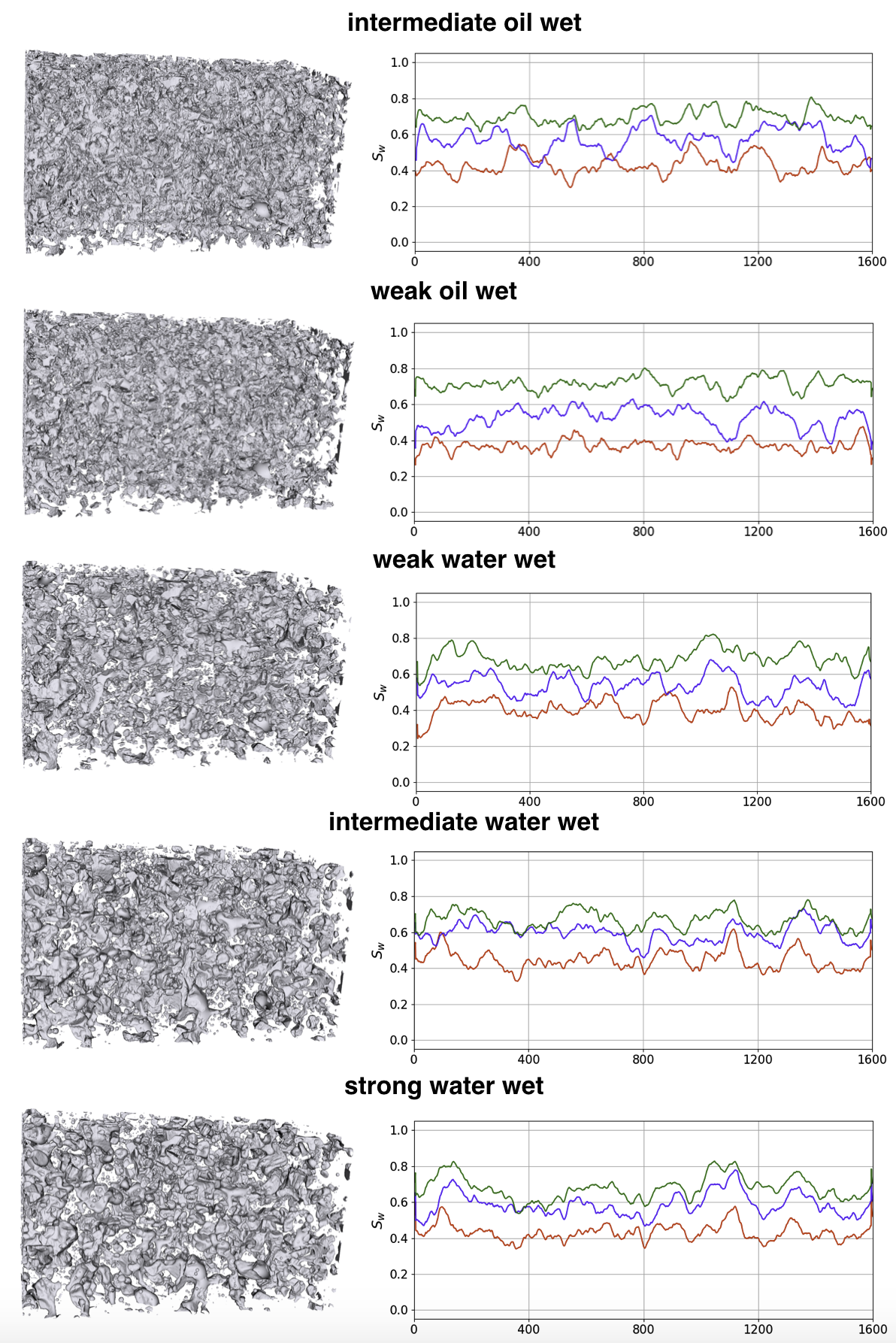}
\caption{ Simulations of steady state relative permeability using periodic boundary condition: (left column) representative oil distribution where water and solid matrix are made transparent; (right column) local water saturation profiles along the flow axis based on for three representative global water saturation. While large saturation gradients are present in water-flooding simulations, nearly uniform saturation profiles with minimal end effects are achieved from steady-state simulation based on the shell aggregation protocol. }
\label{fig:saturation-profile-steady}       
\end{figure*}

Representative local saturation profiles extracted from steady state simulations are shown in Fig.\ref{fig:saturation-profile-steady}. Comparing to the unsteady saturation profiles in Fig.\ref{fig:saturation-profile-unsteady}, a key advantage of the morphological protocols is that boundary effects are much less evident. Most of this is attributable to the morphological algorithms, but is also sensitive
to the configuration of the mixing layers condition, e.g. as shown in Fig. \ref{fig:mixing}. The effect of the mixing layer described in \S \ref{sec:steady-state-protocol} is to increase
the porosity within the mixing layers such that flow pathways will not be blocked by 
the boundary mismatch. This means that the porosity of the mixing region will tend to 
be slightly higher than the porosity of the rest of the sample, with the observed deviation depending on the particular system. However, the effect on the fluid saturation is comparably small. While the Bentheimer sandstone used in this work is 
a relatively open rock, the potential advantage for this condition is even greater for
lower porosity rocks, since the boundary mismatch will tend to block an increasing fraction of the flow pathways as the porosity of the overall system decreases. 

While the saturation gradients during water-flooding are much less significant after breakthrough, this trend presents a barrier to obtaining steady-state relative permeability using the {\em image sequence} protocol, where water-flooding simulation data is used as input. Steady-state configurations cannot be obtained until the saturation gradients are removed. This poses a computationally demanding task, especially at low capillary numbers, where very large numbers of timesteps are required to allow the system to fully relax. For the system considered here at $\mbox{Ca}=10^{-5}$, $\mathcal{O} (10^7)$ timesteps would be required to obtain steady-state flow fields per simulated point. Overall, the required number of timesteps increases proportional to the domain length and is inversely related to the capillary number. On the other hand, morphological based protocol can achieve steady-state flow
an order of magnitude faster, in $\mathcal{O} (10^6)$ timesteps per simulated point, which is due to the fact that the interfaces do not have to move as great a distance.

\begin{figure*}
  \includegraphics[width=1.0\textwidth]{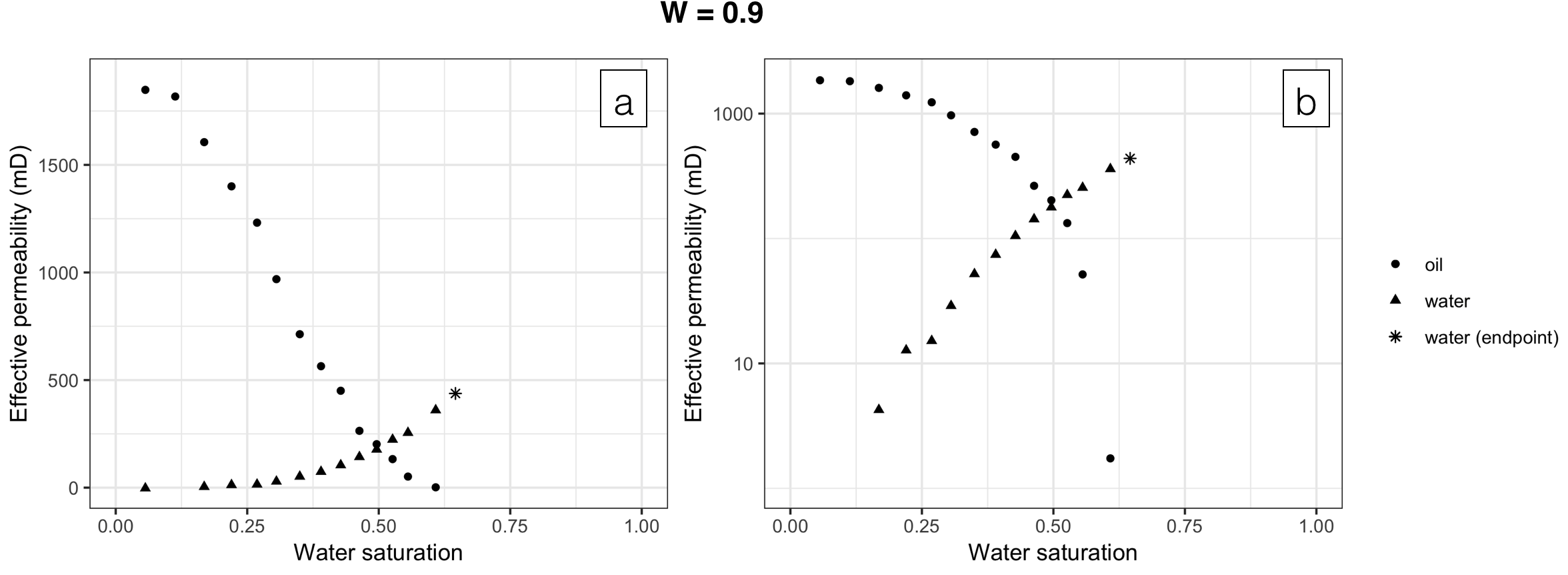}
  \includegraphics[width=1.0\textwidth]{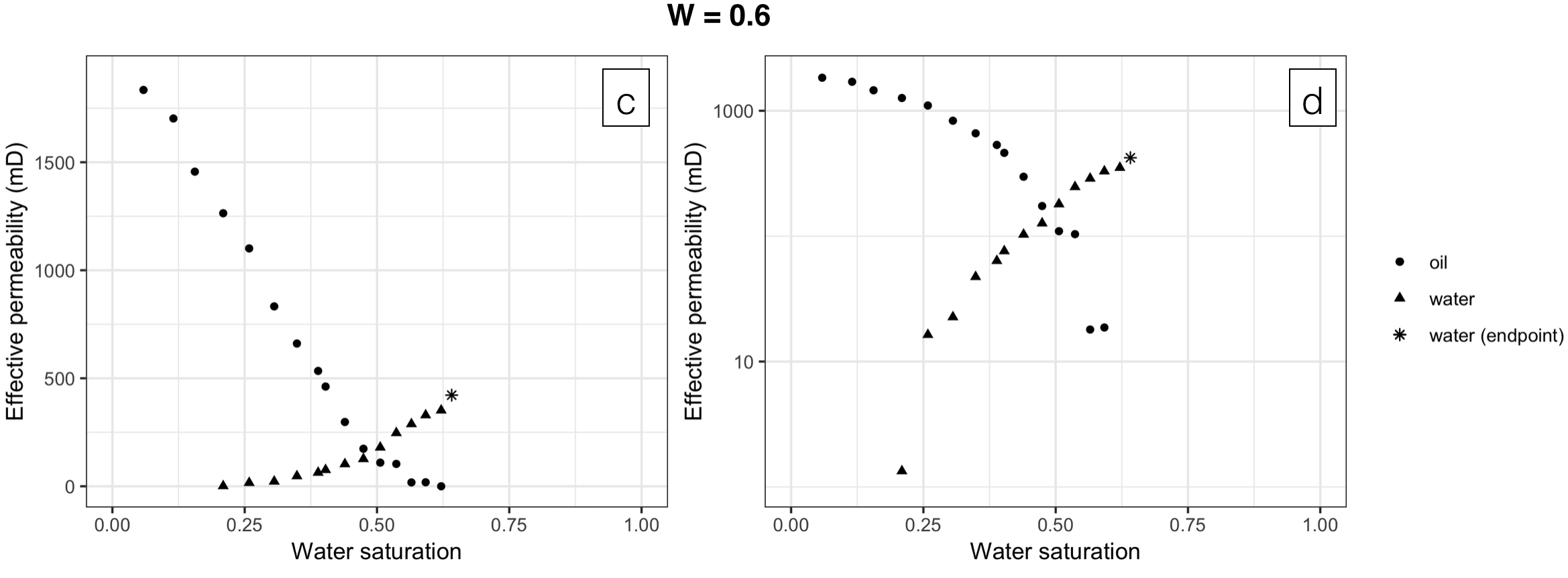}
  \includegraphics[width=1.0\textwidth]{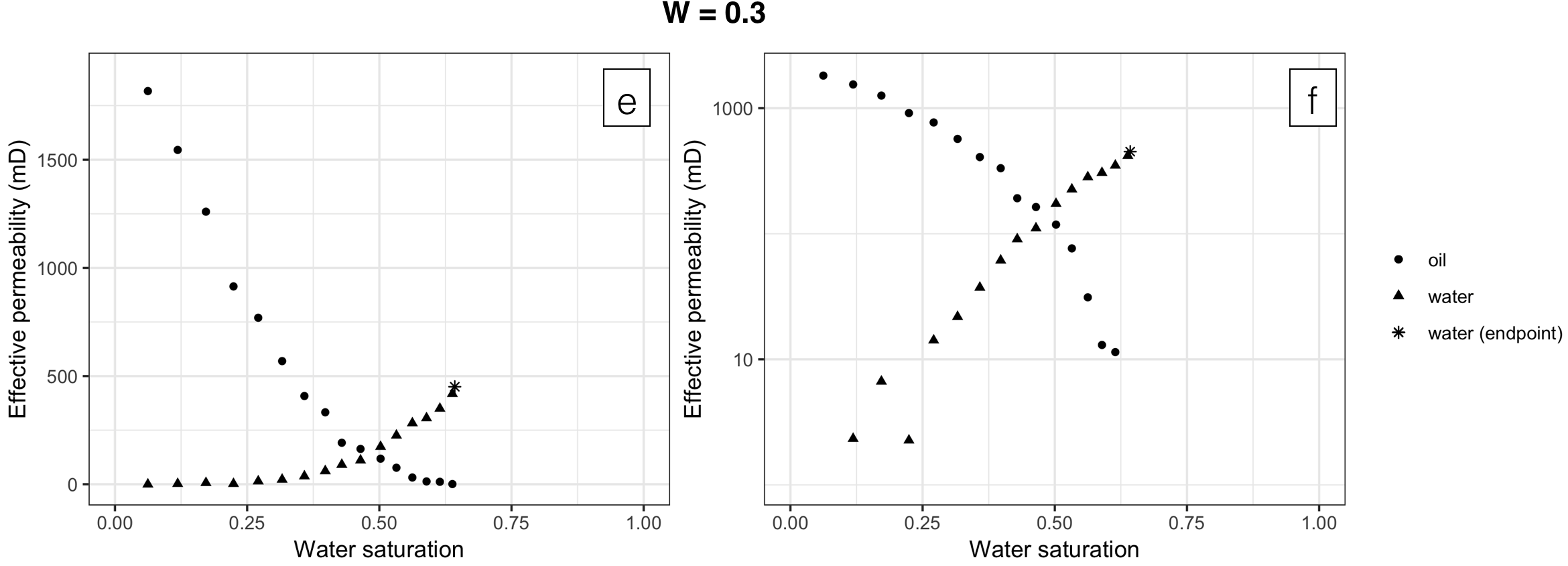}
  \includegraphics[width=1.0\textwidth]{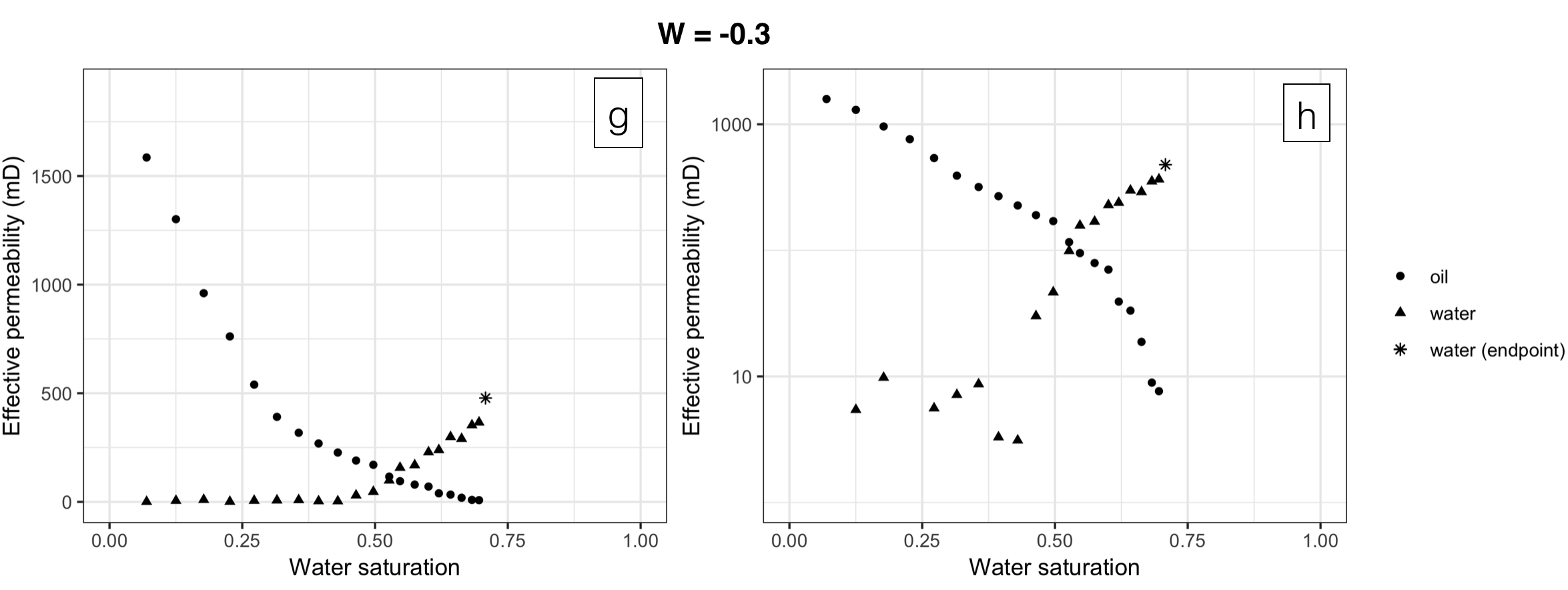}
  \includegraphics[width=1.0\textwidth]{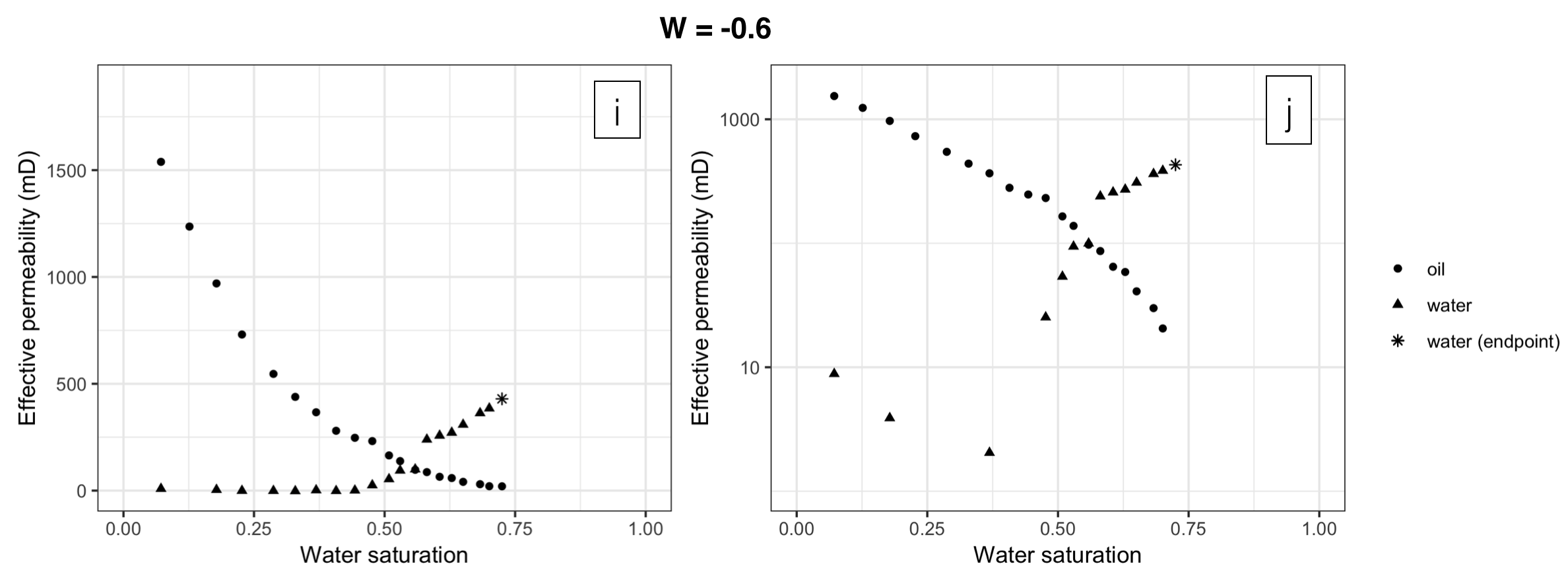}
\caption{Results obtained from simulation of steady-state flow in 
Bentheimer sandstone for various wetting conditions
(a-b) $W=0.9$; (c-d) $W=0.6$; (e-f) $W=0.3$; (g-h) $W=-0.3$; (i-j) $W=-0.6$. The endpoint of water relative permeability is highlighted in asterisk symbol. 
}
\label{fig:kr-sw}       
\end{figure*}

Relative permeability measurements obtained based on steady-state simulations are presented for all wetting conditions in Fig. \ref{fig:kr-sw}. The simulated curves demonstrate expected qualitative behavior for water flooding in mixed wet systems:
(1) water-wet curves exhibit a clear S-shape due to the fact that oil dominates the main flow pathways at low water saturation;
(2) the effective permeability of oil drops more quickly as water saturation 
increases as the wetting condition becomes more oil-wet
(3) the residual oil endpoint shifts based on the wettability, with less trapped oil in more oil-wet conditions;
(4) the effective permeability of water rises gradually due to the swelling of
water films under water-wet conditions;
(5) under more oil-wet conditions, the effective permeability of oil
rises sharply once the percolation threshold is reached. At low water
saturations, the water is not well-connected and cannot flow efficiently.

\begin{figure*}
  \includegraphics[width=1.0\textwidth]{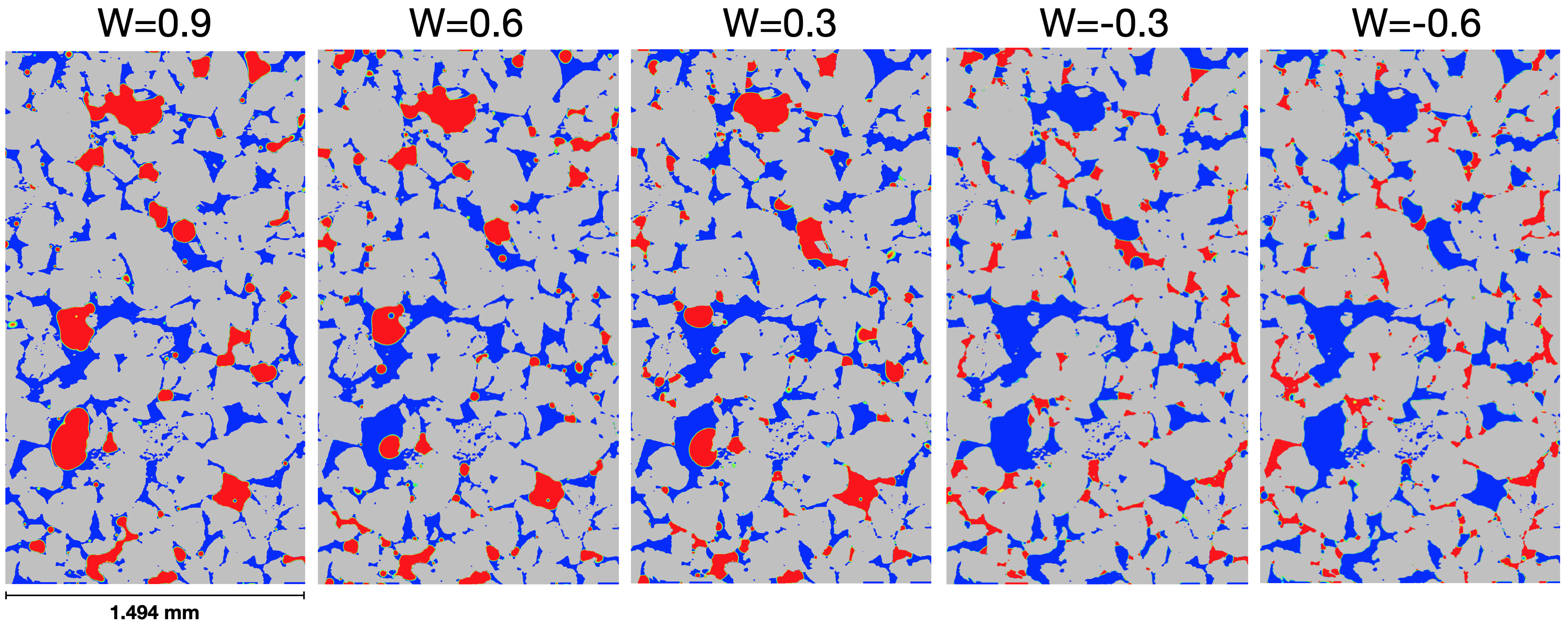}
\caption{Fluid distributions at the oil endpoint based on steady-state simulations
performed using LBPM. The oil (red) will tend to be trapped in the larger pores
under more water-wet conditions and trapped in the tighter corners under
more oil-wet conditions.
}
\label{fig:mutiwet}       
\end{figure*}

Differences in the effective permeability are attributed to the pore-scale distribution
of fluids. In particular, the flow behavior is sensitive to the shape of the connected
pathways that support flow of oil and water through the system. The distribution of
fluids is shown in Fig.\ref{fig:mutiwet} for each wetting condition based on the 
simulated oil endpoint. When conditions are more water-wet, the oil is trapped in
large pores and immobile based on the capillary forces. For more oil-wet conditions,
the oil is trapped in tighter corners, and may flow at a slow rate if  
connected films are formed along the solid grain surface. The distribution of fluids
is different depending on the particular wetting condition, which is the underlying 
cause for differences in the measured effective permeability.

\begin{figure}
\includegraphics[width=1.0\textwidth]{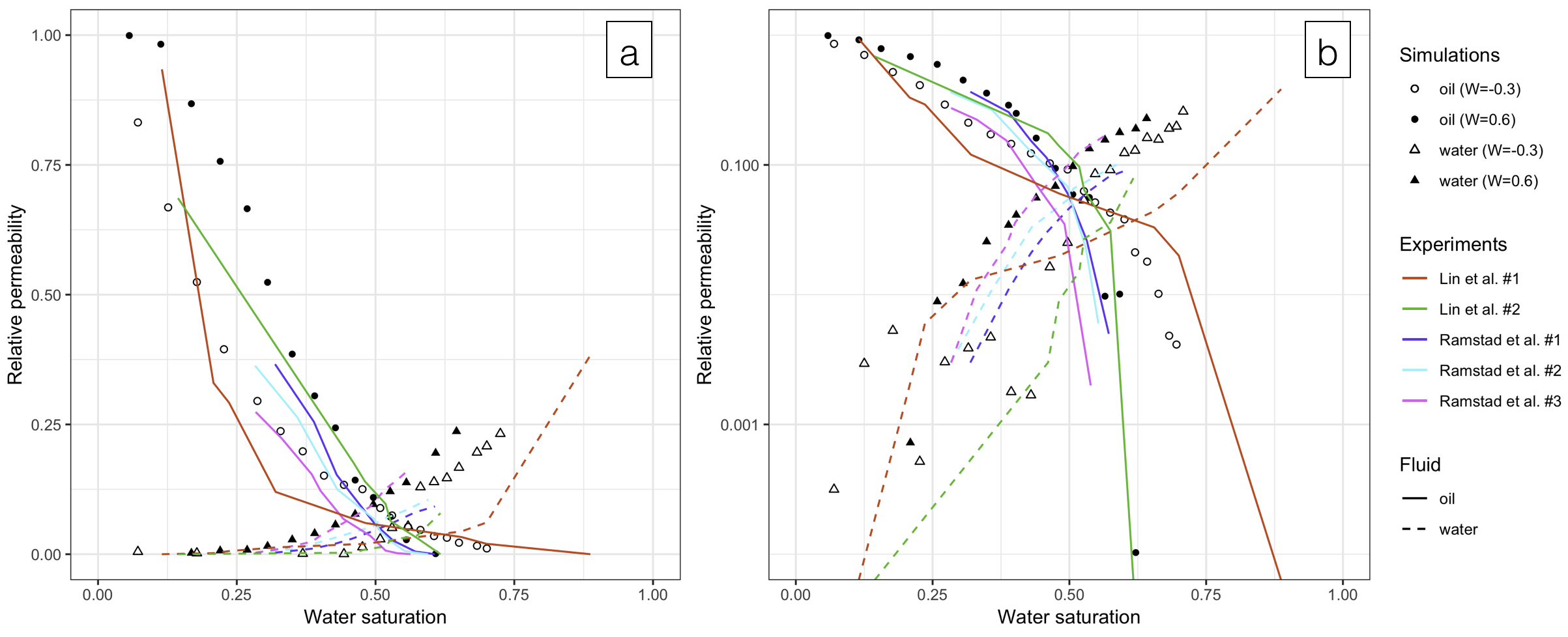}
\caption{Comparison between simulated relative permeability and 
 experimental results reported for two-fluid flow in Bentheimer sandstone, in (a) normal scale, and (b) log scale. }
\label{fig:kr-sw-exp}
\end{figure}

The simulated relative permeability curves are also compared to experimental
results reported in the literature, shown in Fig. \ref{fig:kr-sw-exp}. 
Key information of experimental relative permeability curves for Bentheimer sandstone are listed in Table \ref{tab:kr-sw-exp}.
While contact angles were not directly measured for experiments reported by Ramstad et al. \cite{Ramstad_Idowu_12},
values reported elsewhere in the literature are presumed to be representative. 
For a mixed-wet experiment, Lin at al. report mean contact angle $80^{\circ}$ with standard deviation $17^{\circ}$ (Lin et al. \#1) \cite{PhysRevE.99.063105}.
Under more water-wet conditions (Lin et al. \#2), 
the measured mean contact angle was $66.4^{\circ}$ with
standard deviation $15.1^{\circ}$ \cite{doi:10.1029/2018WR023214}.
A relatively broad distribution of contact angles is typical in real 
porous media; Sun et al. also report experimentally-measured contact angles for oil-water
systems in Bentheimer sandstone, observing a mean contact angle
$58^{\circ}$ with standard deviation $19^{\circ}$ \cite{SUN2020106}. 
Simulated points are plotted on top of the experimental data. Due to the wide range
of contact angles present in natural systems, we show simulated points for the most
strongly water-wet ($\mbox{W=0.9})$ and most strongly oil-wet ($\mbox{W=-0.6}$) cases.
Each of the experimental curves was obtained for a different physical sample, so the 
variations in the relative permeability curves should be partially attributed to 
heterogeneity inherent in the rock structure. Heterogeneity in the rock structure
does not impact the simulated curves. Since the simulated curves are generated 
based on identical rock micro-structure, the wetting condition is solely responsible for the
shift in the relative permeability. It is reasonable to suppose that the wetting conditions
also varied for different experimental samples, even for those cases where contact
angles were not directly measured. An essential challenge with the interpretation of the
experimentally-measured contact angles is that the surface wetting properties can only
be inferred in the immediate vicinity of the contact line. In other words, it should
not be expected that the distribution of the observed contact angles will match the distribution of $\mbox{W}$ along the surface. Since the physics of the system evolve
to move the contact line to a minimum energy configuration, it would be unsurprising if experimentally-measured contact angles deviate from average surface energies in systems where the the surface wetting condition is heterogeneous.

\begin{table}
\caption{Properties for experimental relative permeability curves in Bentheimer sandstone}
\label{tab:kr-sw-exp}       
\begin{tabular}{lcl}
\hline\noalign{\smallskip}
Label & Porosity & Permeability (mD) \\
\noalign{\smallskip}\hline\noalign{\smallskip}
Ramstad et al. \# 1 & $0.241$ & 2820 \\
Ramstad et al. \# 2 & $0.232$ & 2840 \\
Ramstad et al. \# 3 & $0.237$ & 2930 \\
Lin et al. \# 1 & $0.240$ & 2199 \\
Lin et al. \# 2 & $0.240$ & 1490 \\
\noalign{\smallskip}\hline
\end{tabular}
\end{table}

\begin{figure*}
  \includegraphics[width=1.0\textwidth]{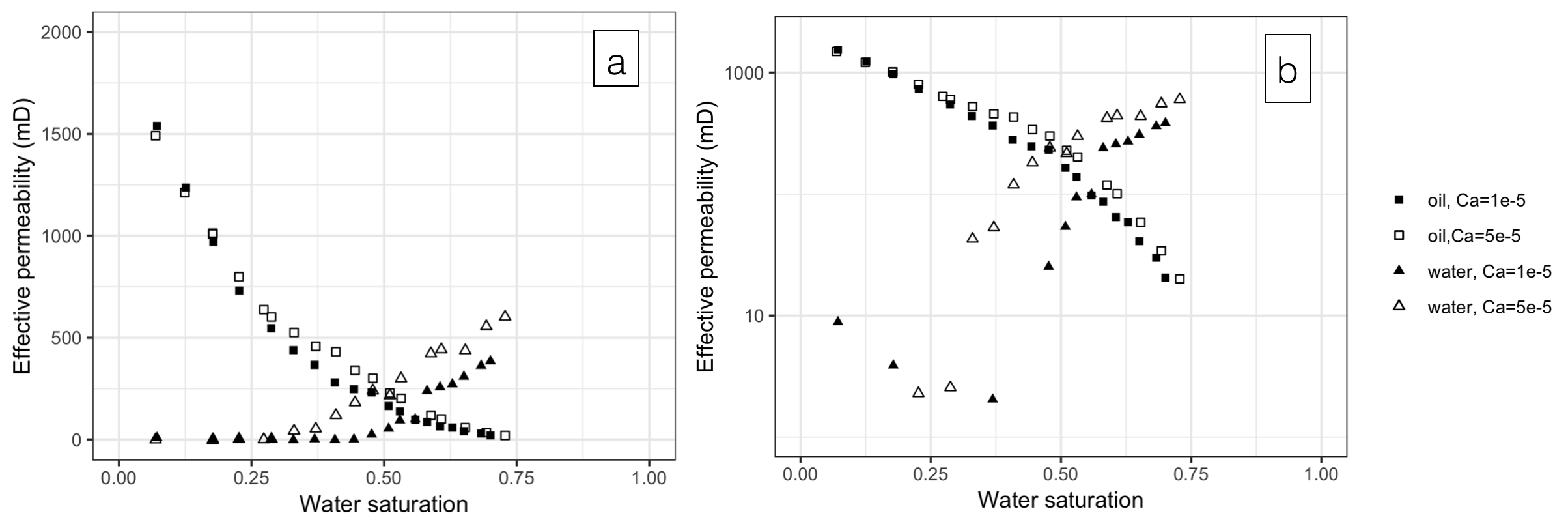}
\caption{ Influence of capillary number on effective permeability 
for intermediate oil wetting condition (W=-0.6).}
\label{fig:kr-sw-Ca}       
\end{figure*}

Capillary number effects influence the overall shape of the effective permeability curves, and have a pronounced effect on the effective permeability of water under more oil-wet conditions. In these situations, 
water flooding corresponds to a drainage process where water displaces oil
from the large pores in the system. Since the large pores hold a larger
volume of water as compared to the films in a water-wet system, more water must be added before the water becomes well-connected.
The influence of capillary number on the effective permeability is shown in 
Fig. \ref{fig:kr-sw-Ca}. 
Here the target capillary number was increased to $\mbox{Ca}=5\times10^{-5}$
for the most strongly oil-wet condition $\mbox{W}=-0.6$. At low water saturation,
the oil effective permeability is slightly reduced, which is possibly due
to inertial effects; eddies are more likely to form when the volume
fraction of one fluid is high. The effective permeability of water rises more quickly
due to the influence of capillary forces on the geometry of the water. Due
to the higher capillary number, water is able to flow more efficiently at lower water saturation due to the fact that viscous forces are strong enough to deform the interfaces and form more efficient flow pathways. At intermediate water saturation, this also has the effect of improving the mobility of oil, which may be due to co-moving interfaces. Furthermore,
the oil endpoint shifts to the right due to viscous mobilization. 

\subsection{Centrifuge protocol for capillary pressure}

\begin{figure*}
  \includegraphics[width=1.0\textwidth]{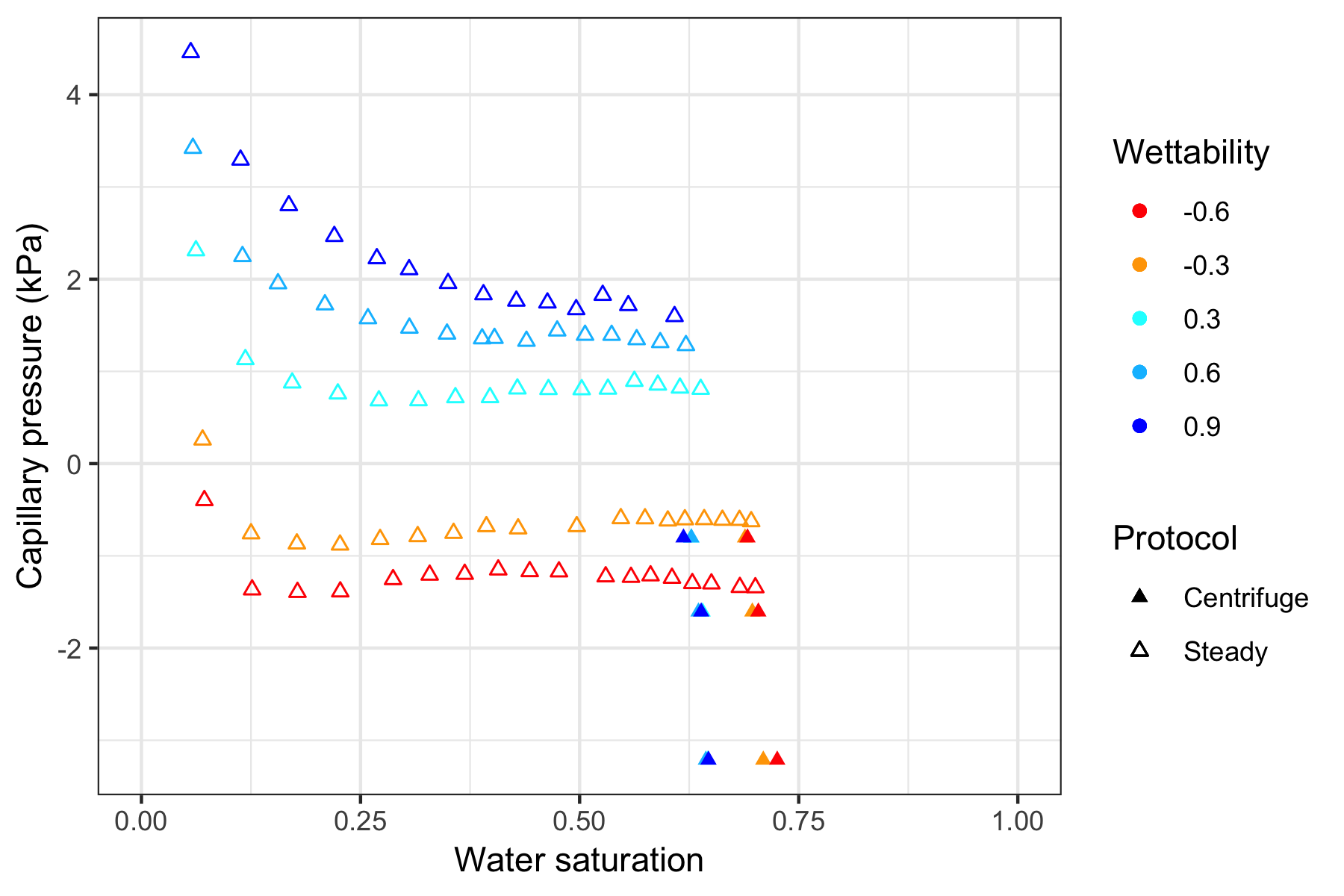}
\caption{Capillary pressure curves obtained from simulation of water flooding in
Bentheimer sandstone. The difference between the average phase pressures can be used to determine the capillary pressure during steady-state simulation. When the endpoint is 
reached, a centrifuge protocol is applied to further assess the endpoint behavior.
}
\label{fig:pc-sw}       
\end{figure*}

Capillary pressure curves can be obtained by applying the centrifuge approach to fluid configurations
produced from steady-state simulations. As shown in Fig. \ref{fig:pc-sw}, the capillary pressure
can be inferred based on the difference in the fluid pressures measured during steady-state flow.
This is obtained directly from volume averaging as described in \S \ref{sec:analysis}. For the steady-state
flow, relatively constant capillary pressures are observed for most saturation values, with the overall wettability of the system shifting the curves up and down. Here the centrifuge approach is initialized
based on the final fluid configuration from the steady-state protocol. Reservoirs for each fluid
are established at opposite ends of the domain (as in Fig. \ref{fig:SCAL_workflow}a) and a body force is
imposed on the system to push more water into the system. Water will displace oil until the 
capillary forces in the system are sufficiently large to block further oil mobilization. When this occurs,
the fluid saturation stabilizes and the associated capillary pressure is inferred from the magnitude of the 
body force, consistent with the interpretation of centrifuge data. Each time the system stabilizes, a new point is obtained along the capillary pressure curve. Additional points are generated by repeating
the procedure after increasing the magnitude of the body force. Depending on the orientation of the body force,
the approach can be used to assess the capillary pressure behavior at either endpoint.

\section{Summary}

We have presented a full computational workflow for digital rock analysis
based on two-fluid flow simulation with the LBPM software. Computational protocols
were developed to mimic SCAL routines, including unsteady core flooding experiments,
steady-state fractional flow experiments used to infer relative permeability, and centrifuge
experiments used to infer capillary pressure. The associated simulation protocols
allow the upscaling of multiphase flow properties from digital images of rock microstructure,
which can be used to inform reservoir simulation. Unsteady simulations were performed using 
conventional pressure and volumetric flux boundary conditions, and steady-state simulations
were performed using fully periodic boundary conditions.
A two-fluid color lattice Boltzmann method was combined with morphological routines to
create an efficient mechanism to update the fluid distribution and change fluid
saturation during steady state simulation. Simulations were performed to assess the 
flow behavior for a wide range of wetting conditions in Bentheimer sandstone. Initial 
configurations were established based on the morphological drainage algorithm,
which was constructed to mimic mercury injection capillary pressure (MICP) in a strongly water-wet
system, which was also used to assess the image resolution. 
Based on the initial configuration, the surface wetting properties were
distributed based on fluid history, with regions of initial oil-contact
having greater affinity toward oil. Unsteady simulations of water-flooding are
shown to recover expected behavior for the displacement, and the associated 
residual oil saturation agrees with expected qualitative trend with respect to 
wettability. Steady-state simulations were used to measure the relative permeability
and compared to previously reported experimental data for Bentheimer sandstone.
Relative permeability trends demonstrate the effect of changes to surface wetting
properties in situations with identical solid micro-structure. Qualitative trends
reported from experiments are also recovered based on simulations. These validations
are compelling evidence that pore-scale simulation can be used to make reliable
inferences on the role of wetting in multiphase flow based on real digital rock data. 
Many unanswered questions remain pertaining to the role of surface wetting phenomena
on larger-scale transport. The numerical implementations described in this work
are freely available through the Open Porous Project under an open source license,
and can be used to support a wide range of simulation studies within the scientific
community.

\begin{acknowledgements}
J.M. and Z.L. thank Equinor ASA for funding parts of this research through a post-doc project. T.R. thanks Anders Kristoffersen, Einar Ebeltoft, Knut Uleberg and {\AA}smund Haugen, Equinor, for valuable discussions. An award of computer time was provided by the Department of Energy Director's Discretionary program and the Frontier Center for Accelerated Application Readiness (CAAR). This research also used resources of the Oak Ridge Leadership Computing Facility, which is a DOE Office of Science User Facility supported under Contract DE-AC05-00OR22725. T.R. acknowledges Equinor ASA for granting permission to publish this paper.
\end{acknowledgements}

%
%

\bibliographystyle{spphys}       


\begin{thebibliography}{100}
\providecommand{\url}[1]{{#1}}
\providecommand{\urlprefix}{URL }
\expandafter\ifx\csname urlstyle\endcsname\relax
  \providecommand{\doi}[1]{DOI \discretionary{}{}{}#1}\else
  \providecommand{\doi}{DOI \discretionary{}{}{}\begingroup
  \urlstyle{rm}\Url}\fi

\bibitem{OPM_2011}
B.~Flemisch, K.~Flornes, K.A. Lie, A.~Rasmussen, AGU Fall Meeting Abstracts pp.
  L08-- (2011)

\bibitem{ISI:000314105900011}
M.J. Blunt, B.~Bijeljic, H.~Dong, O.~Gharbi, S.~Iglauer, P.~Mostaghimi,
  A.~Paluszny, C.~Pentland, {Advances in Water Resources} \textbf{{51}}, {197}
  ({2013}).
\newblock \doi{{10.1016/j.advwatres.2012.03.003}}

\bibitem{bear1988dynamics}
J.~Bear, \emph{Dynamics of Fluids in Porous Media}.
\newblock Dover Civil and Mechanical Engineering Series (Dover, 1988).
\newblock \urlprefix\url{https://books.google.com/books?id=lurrmlFGhTEC}

\bibitem{doi:10.1029/WR012i003p00513}
Y.~Mualem, Water Resources Research \textbf{12}(3), 513 (1976).
\newblock \doi{10.1029/WR012i003p00513}.
\newblock
  \urlprefix\url{https://agupubs.onlinelibrary.wiley.com/doi/abs/10.1029/WR012i003p00513}

\bibitem{opm_lbpm}
J.~McClure.
\newblock {LBPM Software Package} (2020).
\newblock \urlprefix\url{https://github.com/opm/lbpm}

\bibitem{Dalton_2019}
L.~Dalton.
\newblock Bentheimer and nugget residual saturation micro-computed tomography
  data.
\newblock \url{http://www.digitalrocksportal.org/projects/218} (2019).
\newblock \doi{doi:10.17612/P73H4B}

\bibitem{Wildenschild_Sheppard_13}
D.~Wildenschild, A.P. Sheppard, {ADVANCES IN WATER RESOURCES} \textbf{{51}},
  {217} ({2013}).
\newblock \doi{{10.1016/j.advwatres.2012.07.018}}

\bibitem{darcy1856fontaines}
H.~Darcy, \emph{Les fontaines publiques de la ville de Dijon. Exposition et
  application des principes {\`a} suivre et des formules {\`a} employer dans
  les questions de distribution d'eau: ouvrage termin{\'e} par un appendice
  relatif aux fournitures d'eau de plusieurs villes au filtrage des eaux et
  {\`a} la fabrication des tuyaux de fonte, de plomb, de tole et de bitume}.
\newblock Les fontaines publiques de la ville de Dijon. Exposition et
  application des principes {\`a} suivre et des formules {\`a} employer dans
  les questions de distribution d'eau: ouvrage termin{\'e} par un appendice
  relatif aux fournitures d'eau de plusieurs villes au filtrage des eaux et
  {\`a} la fabrication des tuyaux de fonte, de plomb, de tole et de bitume
  (Dalmont, 1856).
\newblock \urlprefix\url{https://books.google.com/books?id=42EUAAAAQAAJ}

\bibitem{Jadhunandan_Morrow_1995}
P.P. Jadhunandan, N.R. Morrow, {SPE Reservoir Engineering} \textbf{{10}}({1}),
  {22597} ({1995}).
\newblock \doi{{10.2118/22597-PA}}

\bibitem{Ramstad_Idowu_12}
T.~Ramstad, N.~Idowu, C.~Nardi, P.E. Oren, {\it Transport in Porous Media}
  \textbf{{94}}({2, SI}), {487} ({2012}).
\newblock \doi{{10.1007/s11242-011-9877-8}}

\bibitem{Adamson_Gast_97}
A.~Adamson, A.~Gast, \emph{{Physical Chemistry of Surfaces}} (Wiley, 1997)

\bibitem{doi:10.1029/98RG00878}
J.C. Muccino, W.G. Gray, L.A. Ferrand, Reviews of Geophysics \textbf{36}(3),
  401 (1998).
\newblock \doi{10.1029/98RG00878}.
\newblock
  \urlprefix\url{https://agupubs.onlinelibrary.wiley.com/doi/abs/10.1029/98RG00878}

\bibitem{Lenormand1989}
R.~Lenormand, C.~Zarcone, Transport in Porous Media \textbf{4}(6), 599 (1989).
\newblock \doi{10.1007/BF00223630}.
\newblock \urlprefix\url{https://doi.org/10.1007/BF00223630}

\bibitem{Erpelding_Sinha_etal_13}
M.~Erpelding, S.~Sinha, K.T. Tallakstad, A.~Hansen, E.G. Flekkoy, K.J. Maloy,
  {Physical Review E} \textbf{{88}}({5}) ({2013}).
\newblock \doi{{10.1103/PhysRevE.88.053004}}

\bibitem{PhysRevE.87.033012}
A.L. Dye, J.E. McClure, C.T. Miller, W.G. Gray, Phys. Rev. E \textbf{87},
  033012 (2013).
\newblock \doi{10.1103/PhysRevE.87.033012}.
\newblock \urlprefix\url{https://link.aps.org/doi/10.1103/PhysRevE.87.033012}

\bibitem{McClure_2010}
J.~McClure, W.~Gray, C.~Miller, Transport in Porous Media \textbf{84}, 535
  (2010).
\newblock \doi{https://doi.org/10.1007/s11242-009-9518-7}

\bibitem{Latva_Kokko_Rothman_05b}
M.~Latva-Kokko, D.~Rothman, {\it \it Physical Review E} \textbf{{71}}({5, Part
  2}) ({2005}).
\newblock \doi{{10.1103/PhysRevE.71.056702}}

\bibitem{Latva_Kokko_Rothman_07}
M.~Latva-Kokko, D.H. Rothman, {\it Physical Review Letters} \textbf{{98}}({25})
  ({2007}).
\newblock \doi{{10.1103/PhysRevLett.98.254503}}

\bibitem{ISI:A1988N050200002}
N.~Wardlaw, Y.~Li, Transport in Porous Media \textbf{{3}}({1}), {17} ({1988}).
\newblock \doi{{10.1007/BF00222684}}

\bibitem{ISI:A1990DD11300001}
G.~Jerauld, S.~Salter, {Transport in Porous Media} \textbf{{5}}({2}), {103}
  ({1990}).
\newblock \doi{{10.1007/BF00144600}}

\bibitem{ISI:000180779800003}
H.~Vogel, in \emph{Morphology of Condensed Matter: Physics and Geometry of
  Spatially Complex Systems}, \emph{{Lecture Notes in Physics}}, vol. {600},
  ed. by {Mecke, K and Stoyan, D} ({2002}), \emph{{Lecture Notes in Physics}},
  vol. {600}, pp. {75--92}.
\newblock {2nd International Wuppertal Workshop on Statistical Physics and
  Spatial Statistics, Univ. Wuppertal, Wuppertal, Germany, March 05-09, 2001}

\bibitem{Schlueter_Sheppard_etal_14}
S.~Schlueter, A.~Sheppard, K.~Brown, D.~Wildenschild, {Water Resources
  Research} \textbf{{50}}({4}), {3615} ({2014}).
\newblock \doi{{10.1002/2014WR015256}}

\bibitem{ISI:000082526700007}
W.~Lindquist, A.~Venkatarangan, {Physics and Chemistry of the Earth Part
  A-Solid Earth and Geodesy} \textbf{{24}}({7}), {593} ({1999}).
\newblock \doi{{10.1016/S1464-1895(99)00085-X}}

\bibitem{ISI:000089336000016}
W.~Lindquist, A.~Venkatarangan, J.~Dunsmuir, T.~Wong, {Journal of Geophysical
  Research-Solid Earth} \textbf{{105}}({B9}), {21509} ({2000}).
\newblock \doi{{10.1029/2000JB900208}}

\bibitem{ISI:000382680900002}
H.~Liu, Q.~Kang, C.R. Leonardi, S.~Schmieschek, A.~Narvaez, B.D. Jones, J.R.
  Williams, A.J. Valocchi, J.~Harting, {Computational Geosciences}
  \textbf{{20}}({4}), {777} ({2016}).
\newblock \doi{{10.1007/s10596-015-9542-3}}

\bibitem{ISI:000341218200003}
I.~Bondino, G.~Hamon, W.~Kallel, D.~Kachuma, {Petrophysics} \textbf{{54}}({6,
  SI}), {538} ({2013}).
\newblock {SCA Symposium, Aberdeen, SCOTLAND, AUG 27-30, 2012}

\bibitem{Joekar-Niasar_vanDijke_etal_12}
V.~Joekar-Niasar, M.I.J. van Dijke, S.M. Hassanizadeh, {\it Transport in Porous
  Media} \textbf{{94}}({2, SI}), {461} ({2012}).
\newblock \doi{{10.1007/s11242-012-0047-4}}

\bibitem{Geller_Krafczyk_etal_06}
S.~Geller, M.~Krafczyk, J.~Tolke, S.~Turek, J.~Hron, Computers \& Fluids
  \textbf{35}(8-9), 888 (2006)

\bibitem{ISI:000344068000012}
H.~Liu, A.J. Valocchi, C.~Werth, Q.~Kang, M.~Oostrom, {Advances in Water
  Resources} \textbf{{73}}, {144} ({2014}).
\newblock \doi{{10.1016/j.advwatres.2014.07.010}}

\bibitem{ISI:000285177400003}
T.~Ramstad, P.E. Oren, S.~Bakke, {SPE Journal} \textbf{{15}}({4}), {923}
  ({2010}).
\newblock \doi{{10.2118/124617-PA}}

\bibitem{ISI:000246548700006}
A.G. Yiotis, J.~Psihogios, M.E. Kainourgiakis, A.~Papaioannou, A.K. Stubos,
  {Colloids and Surfaces A-Physiochemical and Engineering Aspects}
  \textbf{{300}}({1-2, SI}), {35} ({2007}).
\newblock \doi{{10.1016/j.colsurfa.2006.12.045}}.
\newblock {4th International TRI/Princeton Workshop, Princeton, NJ, JUN 21-23,
  2006}

\bibitem{ISI:000269685800006}
H.~Huang, Z.~Li, S.~Liu, X.y. Lu, {International Journal for Numerical Methods
  in Fluids} \textbf{{61}}({3}), {341} ({2009}).
\newblock \doi{{10.1002/fld.1972}}

\bibitem{ISI:000450094200004}
F.O. Alpak, S.~Berg, I.~Zacharoudiou, {Advances in Water Resources}
  \textbf{{122}}, {49} ({2018}).
\newblock \doi{{10.1016/j.advwatres.2018.09.001}}

\bibitem{ISI:000452345000009}
Y.~Shi, G.H. Tang, {International Journal of Heat and Fluid Flow}
  \textbf{{73}}, {101} ({2018}).
\newblock \doi{{10.1016/j.ijheatfluidflow.2018.07.010}}

\bibitem{Thomas_SCA_2019}
T.~Ramstad, A.~Kristoffersen, E.~Ebeltoft, {2018 Annual Symposium of Society of
  Core Analysis} \textbf{SCA2019-003} (2018)

\bibitem{ISI:000449669500002}
M.~Xu, H.~Liu, {European Physical Journal E} \textbf{{41}}({10}) ({2018}).
\newblock \doi{{10.1140/epje/i2018-11735-3}}

\bibitem{ISI:000432555400012}
Z.~Li, S.~Galindo-Torres, G.~Yan, A.~Scheuermann, L.~Li, {Advances in Water
  Resources} \textbf{{116}}, {153} ({2018}).
\newblock \doi{{10.1016/j.advwatres.2018.04.009}}

\bibitem{ISI:000405674300013}
J.F. Xie, S.~He, Y.Q. Zu, B.~Lamy-Chappuis, B.W.D. Yardley, {Heat and Mass
  Transfer} \textbf{{53}}({8}), {2637} ({2017}).
\newblock \doi{{10.1007/s00231-017-2007-6}}

\bibitem{ISI:000404309200006}
H.~Zhao, Z.~Ning, Q.~Kang, L.~Chen, T.~Zhao, {International Communications in
  Heat and Mass Transfer} \textbf{{85}}, {53} ({2017}).
\newblock \doi{{10.1016/j.icheatmasstransfer.2017.04.020}}

\bibitem{ISI:000394568900021}
S.~Mahmoudi, O.~Mohammadzadeh, A.~Hashemi, S.~Kord, {Journal of Petroleum
  Exploration and Production Technology} \textbf{{7}}({1}), {235} ({2017}).
\newblock \doi{{10.1007/s13202-016-0256-4}}

\bibitem{ISI:000388856300003}
G.~Goel, L.K. Abidoye, B.R. Chahar, D.B. Das, {Environmental Fluid Mechanics}
  \textbf{{16}}({5}), {945} ({2016}).
\newblock \doi{{10.1007/s10652-016-9459-y}}

\bibitem{ISI:000383299300006}
D.~Zhang, K.~Papadikis, S.~Gu, {Advances in Water Resources}
  \textbf{{95}}({SI}), {61} ({2016}).
\newblock \doi{{10.1016/j.advwatres.2015.12.015}}

\bibitem{ISI:000383299300013}
S.N. Apourvari, C.H. Arns, {Advances in Water Resources} \textbf{{95}}({SI}),
  {161} ({2016}).
\newblock \doi{{10.1016/j.advwatres.2015.11.005}}

\bibitem{ISI:000321725100003}
Z.~Dou, Z.F. Zhou, {International Journal of Heat and Fluid Flow}
  \textbf{{42}}, {23} ({2013}).
\newblock \doi{{10.1016/j.ijheatfluidflow.2013.01.020}}

\bibitem{ISI:000307391900003}
T.~Ramstad, N.~Idowu, C.~Nardi, P.E. Oren, {Transport in Porous Media}
  \textbf{{94}}({2, SI}), {487} ({2012}).
\newblock \doi{{10.1007/s11242-011-9877-8}}

\bibitem{ISI:000291065000013}
A.~Ghassemi, A.~Pak, {Journal of Petroleum Science and Engineering}
  \textbf{{77}}({1}), {135} ({2011}).
\newblock \doi{{10.1016/j.petrol.2011.02.007}}

\bibitem{ISI:000275765200034}
L.~Hao, P.~Cheng, {International Journal of Heat and Mass Transfer}
  \textbf{{53}}({9-10}), {1908} ({2010}).
\newblock \doi{{10.1016/j.ijheatmasstransfer.2009.12.066}}

\bibitem{ISI:000270378600021}
H.~Huang, X.y. Lu, {Physics of Fluids} \textbf{{21}}({9}) ({2009}).
\newblock \doi{{10.1063/1.3225144}}

\bibitem{Hussain_Pinczewski_14}
F.~Hussain, W.V. Pinczewski, Y.~Cinar, J.Y. Arns, C.H. Arns, M.L. Turner,
  {Transport in Porous Media} \textbf{{104}}({1}), {91} ({2014}).
\newblock \doi{{10.1007/s11242-014-0322-7}}

\bibitem{Landry_Karpyn_etal_14}
C.J. Landry, Z.T. Karpyn, O.~Ayala, {Water Resources Research}
  \textbf{{50}}({5}), {3672} ({2014}).
\newblock \doi{{10.1002/2013WR015148}}

\bibitem{Navaraez_Zauner_etal_10}
A.~Narvaez, T.~Zauner, F.~Raischel, R.~Hilfer, J.~Harting, {\it Journal of
  Statistical Mechanics-Theory and Experiment}  ({2010}).
\newblock \doi{{10.1088/1742-5468/2010/11/P11026}}

\bibitem{Namgyun_10}
N.~Jeong, {\it Transport in Porous Media} \textbf{{83}}({2}), {271} ({2010}).
\newblock \doi{{10.1007/s11242-009-9438-6}}

\bibitem{Maeir_Bernard_10}
R.S. Maier, R.S. Bernard, {\it Journal of Computational Physics}
  \textbf{{229}}({2}), {233} ({2010}).
\newblock \doi{{10.1016/j.jcp.2009.09.013}}

\bibitem{Toelke_Freudiger_etal_06}
J.~Toelke, S.~Freudiger, M.~Krafczyk, {Computers \& Fluids}
  \textbf{{35}}({8-9}), {820} ({2006}).
\newblock \doi{{10.1016/j.compfluid.2005.08.010}}.
\newblock {1st International Conference for Mesoscopic Methods in Engineering
  and Science (ICMMES), Tech Univ Braunschweig, Braunschweig, GERMANY, JUL 25,
  2004-JUL 30, 2005}

\bibitem{Ahrenholz_Toelke_etal_08}
B.~Ahrenholz, J.~Toelke, P.~Lehmann, A.~Peters, A.~Kaestner, M.~Krafczyk,
  W.~Durner, {Advances in Water Resources} \textbf{{31}}({9}), {1151} ({2008}).
\newblock \doi{{10.1016/j.advwatres.2008.03.009}}

\bibitem{Tolke_Krafczyk_etal_02}
J.~Tolke, M.~Krafczyk, E.~Rank, Journal of Statistical Physics
  \textbf{107}(1-2), 573 (2002)

\bibitem{McClure_Prins_etal_14}
J.~McClure, J.~Prins, C.~Miller, Computer Physics Communications
  \textbf{185}(7), 1865  (2014).
\newblock \doi{https://doi.org/10.1016/j.cpc.2014.03.012}.
\newblock
  \urlprefix\url{http://www.sciencedirect.com/science/article/pii/S0010465514000927}

\bibitem{dHumieres_Ginzburg_etal_2002}
D.~d{\textquoteright}Humi{\`e}res, I.~Ginzburg, M.~Krafczyk, P.~Lallemand, L.S.
  Luo, Philosophical Transactions of the Royal Society of London A:
  Mathematical, Physical and Engineering Sciences \textbf{360}(1792), 437
  (2002).
\newblock \doi{10.1098/rsta.2001.0955}.
\newblock
  \urlprefix\url{http://rsta.royalsocietypublishing.org/content/360/1792/437}

\bibitem{dHumieres_Ginzburg_2003}
I.~Ginzburg, D.~d'Humi\`eres, Phys. Rev. E \textbf{68}, 066614 (2003).
\newblock \doi{10.1103/PhysRevE.68.066614}.
\newblock \urlprefix\url{https://link.aps.org/doi/10.1103/PhysRevE.68.066614}

\bibitem{Lallemand_Luo_00}
P.~Lallemand, L.S. Luo, Physical Review E \textbf{61}(6), 6546 (2000)

\bibitem{Graue_Ferno_12}
A.~Graue, M.A. Ferno, E.~Aspenes, R.~Needham, {Journal of Petroleum Science and
  Engineering} \textbf{{94-95}}, {89} ({2012}).
\newblock \doi{{10.1016/j.petrol.2012.06.020}}

\bibitem{Saraji_Goual_etal_13}
S.~Saraji, L.~Goual, M.~Piri, H.~Plancher, {Langmuir} \textbf{{29}}({23}),
  {6856} ({2013}).
\newblock \doi{{10.1021/la3050863}}

\bibitem{Matthew_Bijelic_etal_14}
M.~Andrew, B.~Bijeljic, M.J. Blunt, {Advances in Water Resources}
  \textbf{{68}}, {24} ({2014}).
\newblock \doi{{10.1016/j.advwatres.2014.02.014}}

\bibitem{Rucker_2020}
M.~R\"{u}cker, W.B. Bartels, T.~Bultreys, M.~Boone, K.~Singh, G.~Garfi,
  A.~Scanziani, C.~Spurin, S.~Yesufu-Rufai, S.~Krevor, M.J. Blunt, O.~Wilson,
  H.~Mahani, V.~Cnudde, P.~Luckham, A.~Georgiadis, S.~Berg, Petrophysics
  \textbf{61}(2), {2020} (2020)

\bibitem{ISI:000428474500035}
J.~Zhao, Q.~Kang, J.~Yao, H.~Viswanathan, R.~Pawar, L.~Zhang, H.~Sun, {Water
  Resources Research} \textbf{{54}}({2}), {1295} ({2018}).
\newblock \doi{{10.1002/2017WR021443}}

\bibitem{ISI:000337672900004}
C.J. Landry, Z.T. Karpyn, O.~Ayala, {Water Resources Research}
  \textbf{{50}}({5}), {3672} ({2014}).
\newblock \doi{{10.1002/2013WR015148}}

\bibitem{Huang_Thorne_etal_07}
H.~Huang, D.T. Thorne, Jr., M.G. Schaap, M.C. Sukop, {\it \it Physical Review
  E} \textbf{{76}}({6, Part 2}) ({2007}).
\newblock \doi{{10.1103/PhysRevE.76.066701}}

\bibitem{Wiklund_Lindstrom_etal_11}
H.S. Wiklund, S.B. Lindstrom, T.~Uesaka, {Computer Physics Communications}
  \textbf{{182}}({10}), {2192} ({2011}).
\newblock \doi{{10.1016/j.cpc.2011.05.019}}

\bibitem{Schmieschek_Harting_11}
S.~Schmieschek, J.~Harting, {Communications in Computational Physics}
  \textbf{{9}}({5}), {1165} ({2011}).
\newblock \doi{{10.4208/cicp.201009.271010s}}.
\newblock {18th International Conference on Discrete Simulation of Fluid
  Mechanics (DSFD), Beijing, PEOPLES R CHINA, JUL 06-10, 2009}

\bibitem{Wolf_dosSantos_etal_09}
F.G. Wolf, L.O.E. dos Santos, P.C. Philippi, {\it Journal of Statistical
  Mechanics-Theory and Experiment}  ({2009}).
\newblock \doi{{10.1088/1742-5468/2009/06/P06008}}

\bibitem{Lee_Lin_08}
T.~Lee, L.~Liu, {\it \it Physical Review E} \textbf{{78}}({1, Part 2})
  ({2008}).
\newblock \doi{{10.1103/PhysRevE.78.017702}}

\bibitem{Benzi_Biferale_etal_06a}
R.~Benzi, L.~Biferale, M.~Sbragaglia, S.~Succi, F.~Toschi, {\it Physical Review
  E} \textbf{74}(2) (2006)

\bibitem{Benzi_Biferale_etal_06c}
R.~Benzi, L.~Biferale, M.~Sbragaglia, S.~Succi, F.~Toschi, Journal of Fluid
  Mechanics \textbf{548}, 257 (2006)

\bibitem{Pomeau_02}
Y.~Pomeau, {Comptes Rendus Mecanique} \textbf{{330}}({3}), {207} ({2002}).
\newblock \doi{{10.1016/S1631-0721(02)01445-6}}

\bibitem{Dhori_Slattery_97}
P.~Dhori, J.~Slattery, {Journal of Non-Newtonian Fluid Mechanics}
  \textbf{{71}}({3}), {197} ({1997}).
\newblock \doi{{10.1016/S0377-0257(97)00007-4}}

\bibitem{Shikhmurzaev_97}
Y.~Shikhmurzaev, {Journal of Fluid Mechanics} \textbf{{334}}, {211} ({1997}).
\newblock \doi{{10.1017/S0022112096004569}}

\bibitem{Brochardwyart_DeGennes_92}
F.~Brochardwyart, P.~DeGennes, {Advances in Colloid and Interface Science}
  \textbf{{39}}, {1} ({1992}).
\newblock \doi{{10.1016/0001-8686(92)80052-Y}}

\bibitem{Seppecher_96}
P.~Seppecher, {International Journal of Engineering Science}
  \textbf{{34}}({9}), {977} ({1996}).
\newblock \doi{{10.1016/0020-7225(95)00141-7}}

\bibitem{Pooley_Kusumaatmaja_etal_08}
C.M. Pooley, H.~Kusumaatmaja, J.M. Yeomans, {\it \it Physical Review E}
  \textbf{{78}}({5, Part 2}) ({2008}).
\newblock \doi{{10.1103/PhysRevE.78.056709}}

\bibitem{Kawasaki_Onishi_etal_08}
A.~Kawasaki, J.~Onishi, Y.~Chen, H.~Ohashi, {\it Computers \& Mathematics with
  Applications} \textbf{{55}}({7}), {1492} ({2008}).
\newblock \doi{{10.1016/j.camwa.2007.08.026}}.
\newblock {2nd International Conference on Mesoscopic Methods in Engineering
  and Science (ICMMES), Hong Kong Polytechn Univ, Hong Kong, PEOPLES R CHINA,
  JUL 25-29, 2005}

\bibitem{Zhao13799}
B.~Zhao, C.W. MacMinn, B.K. Primkulov, Y.~Chen, A.J. Valocchi, J.~Zhao,
  Q.~Kang, K.~Bruning, J.E. McClure, C.T. Miller, A.~Fakhari, D.~Bolster,
  T.~Hiller, M.~Brinkmann, L.~Cueto-Felgueroso, D.A. Cogswell, R.~Verma,
  M.~Prodanovi{\'c}, J.~Maes, S.~Geiger, M.~Vassvik, A.~Hansen, E.~Segre,
  R.~Holtzman, Z.~Yang, C.~Yuan, B.~Chareyre, R.~Juanes, Proceedings of the
  National Academy of Sciences \textbf{116}(28), 13799 (2019).
\newblock \doi{10.1073/pnas.1901619116}.
\newblock \urlprefix\url{https://www.pnas.org/content/116/28/13799}

\bibitem{Latva_Kokko_Rothman_05a}
M.~Latva-Kokko, D.~Rothman, {\it \it Physical Review E} \textbf{{72}}({4, Part
  2}) ({2005}).
\newblock \doi{{10.1103/PhysRevE.72.046701}}

\bibitem{doi:10.1002/fld.4822}
Z.~Li, J.E. McClure, J.~Middleton, T.~Varslot, A.P. Sheppard, International
  Journal for Numerical Methods in Fluids \textbf{n/a}(n/a) (2020).
\newblock \doi{10.1002/fld.4822}

\bibitem{Zhang_Kwok_09}
J.~Zhang, D.Y. Kwok, {\it European Physical Journal-Special Topics}
  \textbf{{171}}, {45} ({2009}).
\newblock \doi{{10.1140/epjst/e2009-01010-2}}.
\newblock {16th Discrete Simulation of Fluid Dynamics International Conference,
  Univ Calgary, Schulich Sch Engn, Banff, CANADA, JUL 23-27, 2007}

\bibitem{Hilpert_Adalsteinsson_06}
D.~Adalsteinsson, M.~Hilpert, Transport in Porous Media \textbf{65}(2), 337
  (2006).
\newblock \doi{10.1007/s11242-005-6091-6}

\bibitem{McClure_Wang_etal_2014}
J.E. McClure, H.~Wang, J.F. Prins, C.T. Miller, W.C. Feng, in \emph{Proceedings
  of the 2014 IEEE 28th International Parallel and Distributed Processing
  Symposium} (IEEE Computer Society, USA, 2014), IPDPS ’14, p. 583–592.
\newblock \doi{10.1109/IPDPS.2014.67}.
\newblock \urlprefix\url{https://doi.org/10.1109/IPDPS.2014.67}

\bibitem{McClure:2016:ASC:3018859.3018862}
J.E. McClure, M.A. Berrill, J.F. Prins, C.T. Miller, in \emph{Proceedings of
  the 2Nd Workshop on In Situ Infrastructures for Enabling Extreme-scale
  Analysis and Visualization} (IEEE Press, Piscataway, NJ, USA, 2016), ISAV
  '16, pp. 12--17.
\newblock \doi{10.1109/ISAV.2016.8}.
\newblock \urlprefix\url{https://doi.org/10.1109/ISAV.2016.8}

\bibitem{Schr_der_Turk_2013}
G.E. Schröder-Turk, W.~Mickel, S.C. Kapfer, F.M. Schaller, B.~Breidenbach,
  D.~Hug, K.~Mecke, New Journal of Physics \textbf{15}(8), 083028 (2013).
\newblock \doi{10.1088/1367-2630/15/8/083028}.
\newblock \urlprefix\url{http://dx.doi.org/10.1088/1367-2630/15/8/083028}

\bibitem{Ginzburg_etal_08}
I.~Ginzburg, F.~Verhaeghe, D.~d'Humieres, Communications in Computational
  Physics \textbf{3}(2), 427 (2008)

\bibitem{Zou_He_1997}
Q.~Zou, X.~He, Physics of Fluids \textbf{9}(6), 1591 (1997).
\newblock \doi{10.1063/1.869307}.
\newblock \urlprefix\url{https://doi.org/10.1063/1.869307}

\bibitem{Li_McClure_fluxb_BC}
J.~McClure, Z.~Li, A.~Sheppard, C.~Miller, {arXiv preprint arXiv:1806.10589}
  (2020)

\bibitem{avraam_payatakes_1995}
D.G. Avraam, A.C. Payatakes, Journal of Fluid Mechanics \textbf{293}, 207–236
  (1995).
\newblock \doi{10.1017/S0022112095001698}

\bibitem{Fredrich_Digiovanni_etal_06}
J.T. Fredrich, A.A. Digiovanni, D.R. Noble, Journal of Geophysical
  Research-Solid Earth \textbf{111}(B3) (2006)

\bibitem{PhysRevE.94.043113}
R.T. Armstrong, J.E. McClure, M.A. Berrill, M.~R\"ucker, S.~Schl\"uter,
  S.~Berg, Phys. Rev. E \textbf{94}, 043113 (2016).
\newblock \doi{10.1103/PhysRevE.94.043113}.
\newblock \urlprefix\url{https://link.aps.org/doi/10.1103/PhysRevE.94.043113}

\bibitem{ISI:000277220600014}
E.S. Boek, M.~Venturoli, {Computers \& Mathematics with Applications}
  \textbf{{59}}({7}), {2305} ({2010}).
\newblock \doi{{10.1016/j.camwa.2009.08.063}}

\bibitem{FAN2019522}
M.~Fan, L.E. Dalton, J.~McClure, N.~Ripepi, E.~Westman, D.~Crandall, C.~Chen,
  Fuel \textbf{252}, 522  (2019).
\newblock \doi{https://doi.org/10.1016/j.fuel.2019.04.098}.
\newblock
  \urlprefix\url{http://www.sciencedirect.com/science/article/pii/S0016236119306635}

\bibitem{WANG2020108966}
Y.~Wang, T.~Chung, R.~Armstrong, J.~McClure, T.~Ramstad, P.~Mostaghimi, Journal
  of Computational Physics \textbf{401}, 108966 (2020).
\newblock \doi{https://doi.org/10.1016/j.jcp.2019.108966}.
\newblock
  \urlprefix\url{http://www.sciencedirect.com/science/article/pii/S0021999119306710}

\bibitem{Pan_Hilpert_Miller_PRE_01}
C.~Pan, M.~Hilpert, C.T. Miller, Physical Review E \textbf{64}, 066702 (2001).
\newblock \doi{10.1103/PhysRevE.64.066702}.
\newblock \urlprefix\url{https://link.aps.org/doi/10.1103/PhysRevE.64.066702}

\bibitem{PhysRevE.99.063105}
Q.~Lin, B.~Bijeljic, S.~Berg, R.~Pini, M.J. Blunt, S.~Krevor, Phys. Rev. E
  \textbf{99}, 063105 (2019).
\newblock \doi{10.1103/PhysRevE.99.063105}.
\newblock \urlprefix\url{https://link.aps.org/doi/10.1103/PhysRevE.99.063105}

\bibitem{doi:10.1029/2018WR023214}
Q.~Lin, B.~Bijeljic, R.~Pini, M.J. Blunt, S.~Krevor, Water Resources Research
  \textbf{54}(9), 7046 (2018).
\newblock \doi{10.1029/2018WR023214}.
\newblock
  \urlprefix\url{https://agupubs.onlinelibrary.wiley.com/doi/abs/10.1029/2018WR023214}

\bibitem{SUN2020106}
C.~Sun, J.E. McClure, P.~Mostaghimi, A.L. Herring, D.E. Meisenheimer,
  D.~Wildenschild, S.~Berg, R.T. Armstrong, Journal of Colloid and Interface
  Science \textbf{578}, 106  (2020).
\newblock \doi{https://doi.org/10.1016/j.jcis.2020.05.076}.
\newblock
  \urlprefix\url{http://www.sciencedirect.com/science/article/pii/S0021979720306822}

\end{thebibliography}

\end{document}